\documentclass[journal]{IEEEtran}
\IEEEoverridecommandlockouts
\usepackage{cite}
\usepackage{amsmath,amssymb,amsfonts}
\usepackage[linesnumbered,ruled]{algorithm2e}
\usepackage{graphicx}
\usepackage{textcomp}
\usepackage{xcolor}
\usepackage{epsfig}
\usepackage{multirow}
\usepackage{subfigure}

\usepackage{svg}

\def\BibTeX{{\rm B\kern-.05em{\sc i\kern-.025em b}\kern-.08em
    T\kern-.1667em\lower.7ex\hbox{E}\kern-.125emX}}
\usepackage{cite}

\SetCommentSty{mycommfont}

\begin{document}
\bstctlcite{IEEEexample:BSTcontrol}
\title{A 71.2-$\mu$W Speech Recognition Accelerator with Recurrent Spiking Neural Network }

\author{Chih-Chyau Yang,~\IEEEmembership{Member,~IEEE},~Tian-Sheuan~Chang,~\IEEEmembership{Senior Member,~IEEE} %
\thanks{This work was supported by the National Science and Technology Council, Taiwan, under Grant 111-2622-8-A49-018-SB, 110-2221-E-A49-148-MY3, and 110-2218-E-A49-015-MBK. 
Chih-Chyau Yang is with the Institute of Electronics, National Yang Ming Chiao Tung University, Taiwan, and the Taiwan Semiconductor Research Institute, Taiwan. (e-mail: ccyang@narlabs.org.tw)
Tian-Sheuan Chang is with the Institute of Electronics, National Yang Ming Chiao Tung University, Taiwan. (e-mail: tschang@nycu.edu.tw)}}
\maketitle

\begin{abstract}

This paper introduces a 71.2-$\mu$W speech recognition accelerator designed for edge devices' real-time applications, emphasizing an ultra low power design. Achieved through algorithm and hardware co-optimizations, we propose a compact recurrent spiking neural network with two recurrent layers, one fully connected layer, and a low time step (1 or 2). The 2.79-MB model undergoes pruning and 4-bit fixed-point quantization, shrinking it by 96.42\% to 0.1 MB. On the hardware front, we take advantage of \textit{mixed-level pruning}, \textit{zero-skipping} and \textit{merged spike} techniques, reducing complexity by 90.49\% to 13.86 MMAC/S. The \textit{parallel time-step execution} addresses inter-time-step data dependencies and enables weight buffer power savings through weight sharing. Capitalizing on the sparse spike activity, an input broadcasting scheme eliminates zero computations, further saving power. Implemented on the TSMC 28-nm process, the design operates in real time at 100 kHz, consuming 71.2 $\mu$W,  surpassing state-of-the-art designs. At 500 MHz, it has 28.41 TOPS/W and 1903.11 GOPS/mm$^2$ in energy and area efficiency, respectively.

\end{abstract}


\IEEEpeerreviewmaketitle

\section{Introduction}

Automatic Speech Recognition (ASR) has rapidly become a predominant human-machine interface in both mobile devices and the Internet of Things (IoT) systems. It enriches user experiences in domains such as indoor device control and real-time voice translation. Such real-time and always-on applications necessitate an ultra low-power design.

To address the aforementioned demands, various designs~\cite{asr-review, ntu-timit, jsscc-lstm-asr, jsscc-denoise-asr2023, attention-chorowski2015, las2016, transformer-vaswani2017attention, transformer-moritz2020streaming, se-asr-2022interactive} have been proposed based on artificial neural networks (ANN). Recurrent neural network (RNN)-based designs~\cite{asr-review, ntu-timit, jsscc-lstm-asr, jsscc-denoise-asr2023} are particularly popular due to their ability to capture long temporal information and their relatively smaller model size. However, these RNN-based designs~\cite{ntu-timit, jsscc-lstm-asr, jsscc-denoise-asr2023} still grapple with high power consumption, primarily due to their multibit MAC (multiplication and accumulation) computational complexity, limiting their ultra low power and real-time deployment on edge devices. The design featuring a binary weight convolutional neural network (BCNN) \cite{tcas-1-bcnn} offers reduced MAC computation complexity and achieves a reduction in the error rate by leveraging a self-learning technique.

To address the need for reduced complexity without sacrificing performance, Spiking Neural Networks (SNNs) \cite{snn-wozniak2020deep, wu2023deep-1, li2023inputaware, wu2020deep-2, spiking-lstm-mutli-bit-2022, spiking-lstm-energyefficient2022} have emerged as an attractive alternative in recent years. These networks generate spike signals only when the membrane potential voltage reaches a predetermined threshold. Benefiting from single-bit spike computation and high signal sparsity, SNNs show promise for ultra low power hardware deployment.

However, current SNN models for ASR do not present a compelling reduction in complexity compared to their ANN counterparts. For example, the study in \cite{wu2020deep-2} utilizes a high-dimensional Multi-Layer Perceptron (MLP) network to achieve superior performance, but this results in an increased model size and computational demand. The recurrent SNN in \cite{spiking-lstm-mutli-bit-2022} leverages the enhanced inherent recurrence dynamics of SNNs to reduce the model size and address the challenge of gradient vanishing, while also improving sequence learning. However, its reliance on multibit output over single-bit spike output diminishes the benefits associated with single-bit spike computation and the innate spike sparsity of SNNs. Another research \cite{spiking-lstm-energyefficient2022} proposes a spiking long short-term memory (LSTM) network, drawing from the ANN-to-SNN conversion methodology. However, this approach only incorporates a subset of activation functions for conversion to integrate-and-fire activations, limiting the exploitation of high spiking sparsity. Additionally, the need for numerous time steps to maintain performance inadvertently introduces significant computational complexity. Furthermore, this design requires accessing weight memory at each time step, resulting in substantial power consumption for memory operations.

Existing SNN accelerators \cite{spiking-od-chang, spinalflow_2020, tcas-ii-snn-dvs} primarily target image applications and are designed to support convolutional neural networks. A notable limitation of these accelerators is their heightened latency, arising from repeated access to the same weight data across multiple time steps, leading to extraneous energy consumption. Furthermore, the hardware implemented for zero-skipping in these sparse designs not only introduces load imbalances among processing elements (PEs) but also results in area overhead due to the need for additional nonzero index buffers.

This paper proposes a sub-milliwatt ultra-low-power speech recognition accelerator through algorithm and hardware co-optimizations. For such low power consumption, the complexity and model size shall be minimized as much as possible. Thus, we first reduce computational complexity by proposing a low time step recurrent spiking neural network (RSNN) for speech recognition. RSNN extends the standard RNN to a spiking version with one or two time steps only. Unlike multibit outputs in \cite{spiking-lstm-mutli-bit-2022}, our model uses spike outputs, eliminating MAC computations and ensuring high spike sparsity. The complexity is further reduced by \textit{mixed-level pruning} and 4-bit quantization to reduce the model size by 96.42\% to 0.1 MB, allowing on-chip operation without external DRAM. Techniques like \textit{mixed-level pruning}, \textit{zero-skipping} and \textit{merged spike} cut computational complexity by 90.49\%. Then, to reduce power consumption in the hardware implementation, our design employs \textit{parallel time steps} to optimize weight sharing, effectively halving memory access requirements and thereby conserving power. Furthermore, the merged spike approach, combined with simple bit-wise \textit{OR} and \textit{AND} logic, significantly reduces the operation cycle count. This reduction, consequently, minimizes the operation frequency, leading to additional power savings. The integration of a zero-skipping mechanism with spike broadcasting prevents load imbalance in PEs and eliminates the overhead associated with nonzero indexing, further reducing power consumption. The finalized design consumes only 71.2 $\mu$W, outperforming existing state-of-the-art solutions.

The rest of the paper is organized as follows: Section II introduces the proposed recurrent SNNs model, the complexity/data flow analysis and design optimization. Section III details the deep-learning accelerator architecture, a simple zero-skipping hardware, and efficient data flows in hardware. Implementation results and design comparisons are given in Section IV. Section V concludes the paper.

\begin{figure}[t]
	\centering{\includegraphics[width=0.48\textwidth]{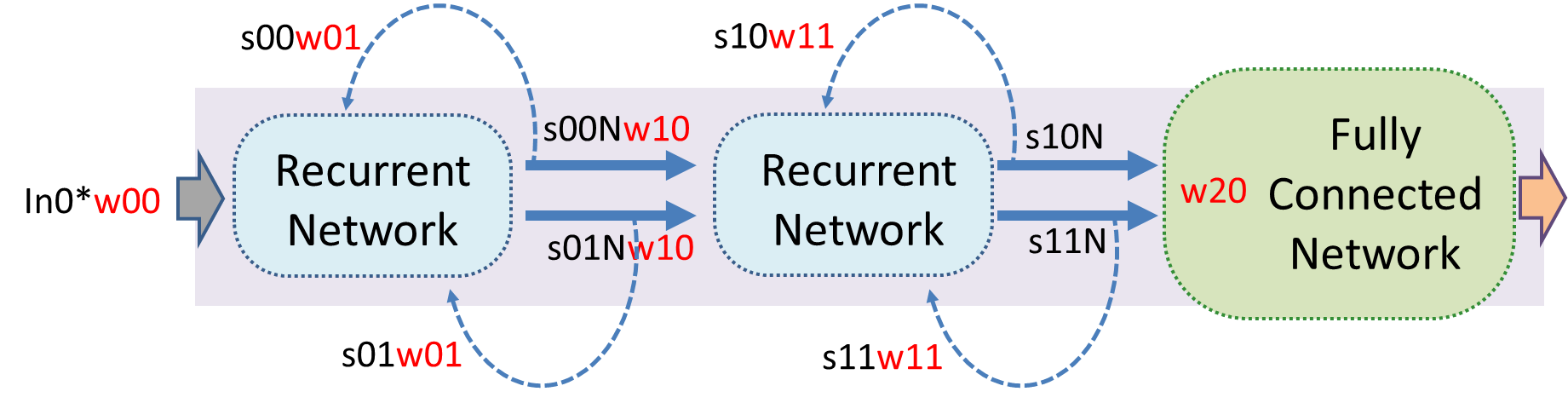}}
	\caption{The proposed RSNN spanning two time steps}
	\label{fig:rnn-2-time-steps}
\end{figure}

\begin{table}[]
    \centering
	\caption{Layer input/output dimensions in baseline and its pruned models}
	\label{table:baseline}
\begin{tabular}{|c|c|c|c|}
\hline
                & \begin{tabular}[c]{@{}c@{}}Baseline \\ model\end{tabular} & \begin{tabular}[c]{@{}c@{}}+ Structured \\ pruning\end{tabular} & \begin{tabular}[c]{@{}c@{}}+ Unstructured \\ pruning\end{tabular} \\ \hline
RNN L0-input        & (40, 256)           & (40, 128)              & (40, 128)           \\ \hline
RNN L0-recurrent    & (256, 256)          & (128, 128)             & (128, 128)          \\ \hline
RNN L1-feedforward  & (256, 256)          & (128, 128)             & (128, 128)          \\ \hline
RNN L1-recurrent    & (256, 256)          & (128, 128)             & (128, 128)          \\ \hline
FC & (256, 1920)         & (128, 1920)            & (128, 1920)         \\ \hline
Parameters      & 698368              & 300032                 & 201728              \\ \hline
\end{tabular}
\end{table}

\section{Complexity/Data Flow analysis and Design Optimization}

\subsection{Proposed Recurrent SNNs Model}

\label{subsection:baseline_model}

Fig.~\ref{fig:rnn-2-time-steps} illustrates the proposed RSNN model for two time steps, acting as the baseline model in this paper. The baseline model features two recurrent layers, each with a dimension of 256, and a fully-connected (FC) layer with a dimension of 1920, representing 698368 parameters. Table~\ref{table:baseline} details the input/output dimensions for layers in both the baseline and its pruned models. The proposed baseline model synergizes the RNN's capability for long temporal information retrieval and compact model size with the SNN's spiking computation and high sparsity attributes.

In the RNN layer, the hidden state $h[t]$ is expressed as $h[t] = f(x[t]W_{x} + h[t-1]W_{h})$, where $f$ represents the activation function. The terms $x[t]$ and $h[t-1]$ denote the input and hidden state at time $t$ and $t-1$, respectively, with $W_{x}$ and $W_{h}$ indicating their respective weights. When considered in conjunction with the SNN operation, the hidden state $h[t][ts]$ is formulated as Eq.~\eqref{eq:1}, where $LIF$ signifies the Leaky Integrate-and-Fire operation and $ts$ denotes the time step in SNN operations. As demonstrated in \cite{snn_lessons, spiking-lstm-energyefficient2022}, the LIF neuron captures computational dynamics to generate spike signals. The dynamics of the LIF neuron at time step $ts$ is described by Eq.~\eqref{eq:2}:
\begin{equation} \label{eq:1}
h[t][ts] = \textit{LIF}(x[t][ts]W_{x} + h[t-1][ts]W_{h}),
\end{equation}
\begin{align}
\label{eq:2}
U[t][ts] &= x[t][ts]W_{x} + h[t-1][ts]W_{h} \notag \\
&\quad + \beta \times U[t][ts-1] \times (1 - h[t][ts-1]),
\end{align}
\noindent where $U[t][ts]$ signifies the membrane potential voltage, $\beta$ stands for the decay factor, and $h[t][ts-1]$ indicates the hidden state at time step $ts-1$. The hidden state or spike output of the RNN layer at time step $ts$, represented by $h[t][ts]$, is determined as:
\begin{equation} \label{eq:3}
h[t][ts] =\begin{cases} 
1 & \text{if } U[t][ts] \geq V_{th}, \\
0 & \text{if } U[t][ts] < V_{th},
\end{cases}
\end{equation}

\noindent with \(V_{th}\) being the threshold potential for the spike signal. To navigate the non-differentiability challenge of spike signals during RSNN training's backpropagation, we utilize a surrogate function~\cite{spiking-lstm-energyefficient2022}, instead of the spike signals in the forward pass.

To enable RSNN training without resorting to multibit outputs as in \cite{spiking-lstm-mutli-bit-2022}, our model incorporates learnable thresholds and decay factors for LIF neurons~\cite{trainable_Vth_D}. Furthermore, to minimize the number of required time steps, we employ the inherent temporal training technique \cite{temporal_inherent_training2021}. This approach begins by training a model with a high time step count and then progressively reduces the time steps, utilizing the higher time step model as a pre-trained model. Based on the experimental results shown in Section IV, we adopt the two time steps for high performance and single time step for low-complexity execution, which offers choices for users.

\begin{figure}[t]
	\centering{\includegraphics[width=0.45\textwidth]{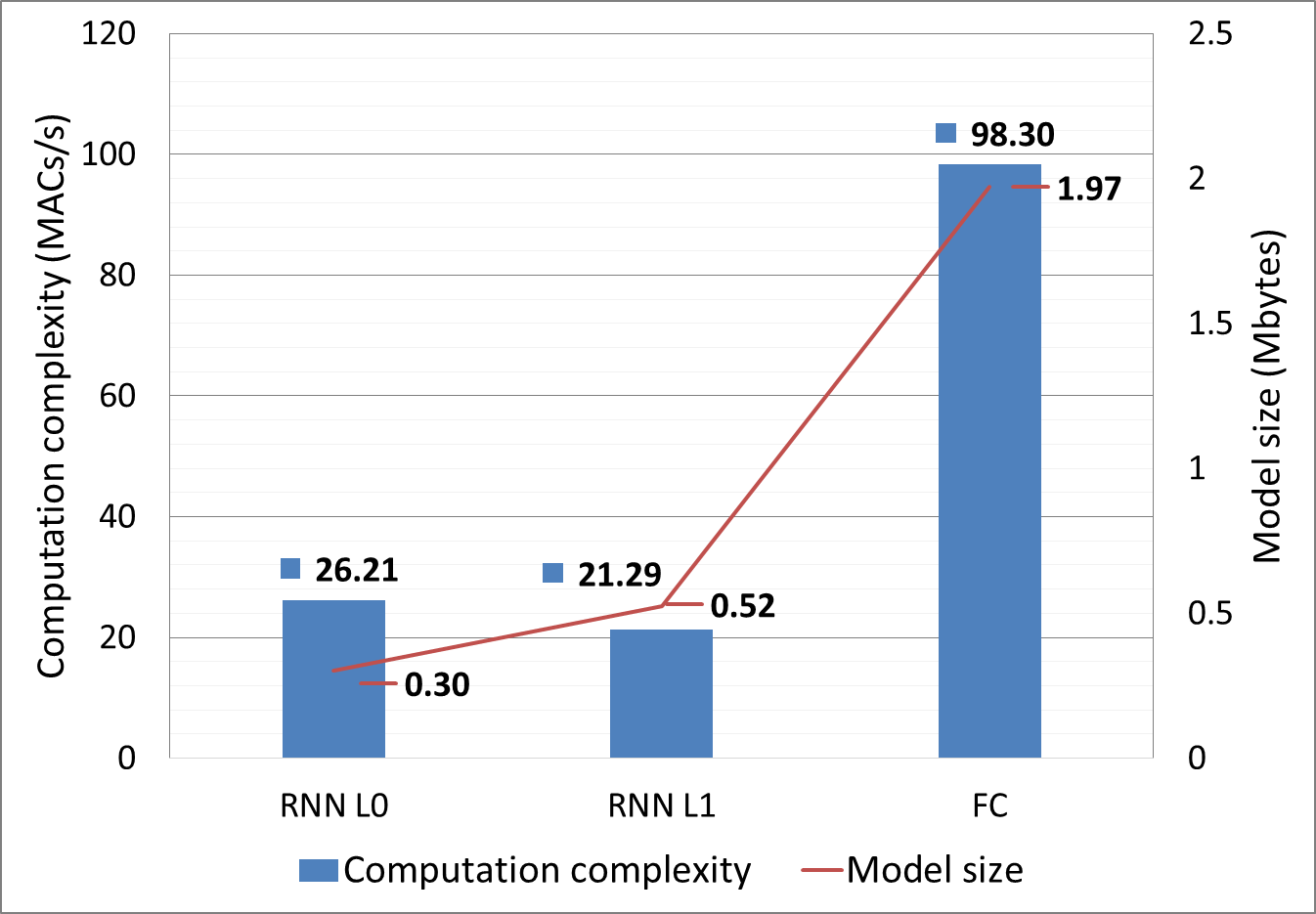}}
	\caption{Computation complexity and weight size of the proposed RSNN model}
	\label{fig:baseline}
\end{figure}

\begin{figure}[t]
   \centering{\includegraphics[width=0.48\textwidth]{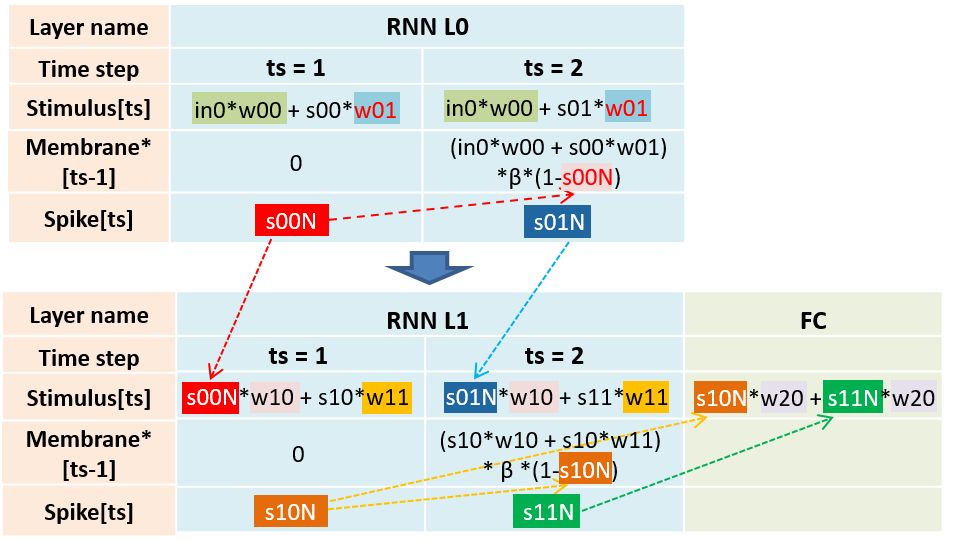}}
	\caption{Data dependencies across time steps and network layers. Note that * indicates that the value of $\text{Membrane}[ts-1]$ is adjusted by $\beta$ and $\text{Spike}[ts-1]$. }
	\label{fig:data_dependency}
\end{figure}

\subsection{Complexity analysis}

To assess the hardware resources necessary for our accelerator, we present a complexity analysis of the proposed model for 1-second speech data in Fig.~\ref{fig:baseline}. The overall model size is 2.79 MB, with the first and second recurrent layers constituting 29.39\% and the fully connected layer making up the remaining 70.61\%. In terms of computational complexity, the model demands 145.8 MMAC/S (Mega Accumulations per second) when processing 25-ms speech features with a 10-ms timing shift. Within this, the first and second recurrent layers account for 32.58\%, while the fully connected layer contributes the remaining 67.42\%. Given such substantial complexity, the task of deploying the model on ultra low power edge devices poses a notable challenge.

\subsection{Data Flow Analysis}
\label{subsection:data_dependency}

Unlike traditional RNN designs, RSNNs pose challenges due to the extra time step dimension. The extra dimension will need to access the weight buffer for each time step, which consumes the most significant power for an accelerator~\cite{spiking-od-chang}. To better understand the data flow, we first investigate the data dependency, as depicted in Fig.~\ref{fig:data_dependency}. In the first recurrent layer, the stimulus signals at \( ts=1 \) and \( ts=2 \) are given by \( in_{0} \times w_{00} + s_{00} \times w_{01} \) and \( in_{0} \times w_{00} + s_{01} \times w_{01} \), respectively. Here, \( in_{0} \) and \( w_{00} \) typify the 40 8-bit input features and their corresponding weight. Symbols \( s_{00} \), \( s_{01} \), and \( w_{01} \) denote spike signals from the recurrent path and their respective weight. Each stimulus signal at \( ts=1 \) and \( ts=2 \) aggregates with the membrane value at time step \( ts-1 \), with the summation sent to the LIF module, yielding spike outputs \( s_{00N} \) and \( s_{01N} \), respectively. This membrane value is defined by Eq.~\eqref{eq:2}. Operations in the subsequent recurrent layer mirror the first, omitted for clarity. In the FC layer, spike signals \( s_{10N} \) and \( s_{11N} \) from the second RNN layer at \( ts=1 \) and \( ts=2 \) are multiplied with weight \( w_{20} \) to compute the final results.

From this analysis, we discern three potential data flows: layer-based, full unfolding, and time step unfolding. The first, a prevalent method, processes the model layer-by-layer, with each time step within the layer executed sequentially. This approach lacks data reuse opportunities for weights and spike activations between layers or time steps, resulting in a weight access count of 1.458 M/s.

The full unfolding data flow concurrently conducts operations across all layers in a pipelined fashion. While supporting \textit{parallel time steps} and weight reuse across time steps, it curtails the model's weight access count to 0.77 M/s. Nevertheless, this method has pitfalls: sparse spike processing can induce load imbalances, undermining hardware utilization. There is also a substantial overhead for the weight buffer, and the critical timing path is restricted to 10-ms throughput.

Conversely, the time step unfolding strategy processes different time steps in tandem, ensuring weights for two time steps are fetched once and broadcast across all time steps. Different layers are handled sequentially. This strategy also yields a weight access count of 0.77 M/s but merely necessitates parallel operations across time steps. Our analysis steers us towards the third dataflow and dual processing element (PE) sets for executing the proposed RSNN model in our accelerator design.

\subsection{Design Optimization}

Given the aforementioned data flow, our design uses \textit{parallel time steps} for the whole model to optimize weight buffer access and employs the \textit{merged spike} technique in the FC layer for complexity reduction. Additionally, to streamline complexity, we apply \textit{mixed-level pruning} to the model and quantize it to a 4-bit fixed point format. This results in minimal performance degradation. Further details are elaborated upon in the following sections.

\subsubsection{Parallel time steps technique}
\label{subsection:weight_reuse}

Memory accesses frequently dominate energy consumption in accelerator designs\cite{sze2017efficient, spiking-od-chang}. In the calculations of our proposed RSNN model, the memory bandwidth required for spike activations is minimal, whereas the demand for weight memory bandwidth is pronounced. Consequently, minimizing weight memory access emerges as a pivotal step in energy reduction. To address this, our paper introduces \textit{parallel time steps}, which decreases the frequency of weight accesses, leading to a more energy-efficient design.

Drawing upon the data dependency analysis detailed in Section \ref{subsection:data_dependency}, we advocate for these parallel time steps. As illustrated in Fig.~\ref{fig:data_dependency}, spike computations in the initial recurrent layer are \( s_{00} \times w_{01} \) at \( ts=1 \) and \( s_{01} \times w_{01} \) at \( ts=2 \). Given that computations in distinct time steps employ the identical weight value \(w_{01}\), this weight can be fetched a single time and subsequently shared between the two spike computations. This strategy is similarly applicable to spike computations in the second recurrent layer; for brevity, we have opted not to delve into it here. Furthermore, for the calculations \( s_{10N} \times w_{20}\) and \(s_{11N} \times w_{20} \) in the FC layer, the weight value \(w_{20}\) can be retrieved just once and then shared between both operations. By harnessing the proposed \textit{parallel time steps} approach, we can slash weight accesses by half.

\subsubsection{Merged spike technique}
\label{subsection:merged_spike}
The number of computation cycles for a real-time speech recognition task determines the minimum operation frequency required for the accelerator. Based on the data dependency analysis in Section \ref{subsection:data_dependency}, we propose the \textit{merged spike} technique for the FC layer to reduce cycle count. As illustrated in Fig.~\ref{fig:data_dependency}, the two spike operations in the FC layer, \( s_{10N} \times w_{20} \) and \( s_{11N} \times w_{20} \), utilize the same weight value \(w_{20}\) and are subsequently summed to obtain the final results. Thus, we merge these two operations: the two spike signals are first summed, and then the multiplication involving the merged spike and the weight is implemented with shift and add operations using the existing hardware. Specifically, a bit-wise \textit{AND} logic determines whether to shift left by a value of 0 or 1, while a bit-wise \textit{OR} logic determines whether the merged spike is zero or not. If it is a zero merged spike, such operations are skipped using the proposed zero-skipping hardware. Otherwise, weight accumulation or weight shift-left accumulation is carried out based on the result of the bit-wise \textit{AND} operation. Employing the \textit{merged spike} technique, the cycle count for merged spike operations can be reduced by 50\%.

\subsubsection{Model compression}
\label{subsubsection:model_compression}

To further reduce complexity, we adopt mixed structured and unstructured pruning on model weights to maximize the possible benefits. For the structured pruning, we adopt the predefined structured pruning technique \cite{2019rethinkingpruning}, which directly shrinks the model structure to the target size and then trains it from scratch. In this paper, we gradually reduce the channel width while maintaining the dimensions of the FC layer to fit the requirements of the decoding module, until both the target size and accuracy constraints are met. The pruned network also serves as the initial input for the unstructured pruning algorithm. In the case of the unstructured pruning, our unstructured pruning flow employs the approach presented in \cite{speechcompression}. Unlike structured pruning, which maintains a regular network structure, unstructured pruning results in irregular weight distribution.
Beyond pruning, we also quantize model weights with quantization-aware training method \cite{quant-aware-training} since activation is already spike signals. Note that the input features are also quantized to 8-bit fixed-point format by examining its dynamic range and impact on performance.

\begin{figure}[t]
	\centering{\includegraphics[width=0.45\textwidth]{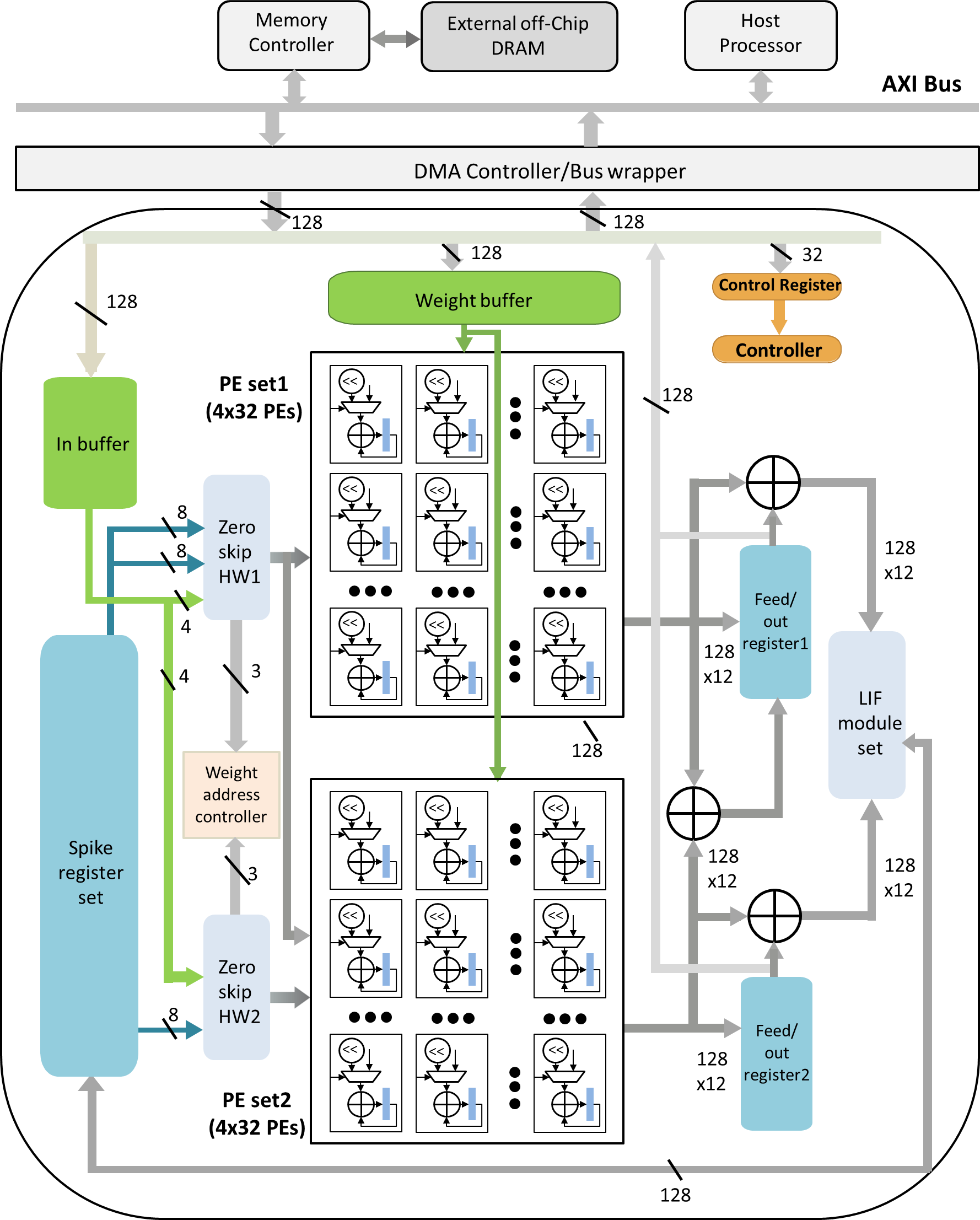}}
	\caption{System architecture}
	\label{fig:system-architecture}
\end{figure}
\section{System Architecture}
\subsection{Overview}

Fig.~\ref{fig:system-architecture} shows the accelerator architecture for speech recognition based on the \textit{parallel time steps} to maximize weight data reuse.  It comprises two sets of 128 parallel 12-bit PEs for two time steps. Each PE is just an accumulator that accumulates \textit{AND} results of the spike input and weight. The PE input includes a 3-bit shifter to shift weights for the input and FC layers. All network weights are loaded into weight buffers (one 48$\times$512-bit and two 192$\times$512-bit weight buffers for the recurrent layers and two 960$\times$512-bit weight buffers for the FC layer) during initialization, eliminating the need for repeated accesses from external DRAM during the model execution, and thus conserving power. The input is loaded into a 48$\times$8-bit input buffer, leveraging interleaved computations between different layers.
However, beyond that, this design shall also need to support single time-step execution and exploit high sparsity of SNN for low power. 
The single time-step execution is easily achieved by reconfiguring PEs. For the high sparsity of SNN, this paper proposed a reconfigurable zero-skipping for different workloads. The detailed data flows are shown below.

\subsection{Reconfigurable Zero Skipping with Input Broadcasting}

Skipping computations for zero inputs and weights are widely used in deep learning accelerators to speed up execution~\cite{nctulaisparsecnn, dateloadbalance, hansoneie, cnvlutin, spiking-od-chang, tcas-ii-snn-dvs}. The compressed RSNN model also has high sparsity due to sparse spiking and unstructured weight pruning. However, previous zero-skipping schemes encounter two problems in hardware design: load imbalance \cite{dateloadbalance, nctulaisparsecnn, spiking-od-chang, tcas-ii-snn-dvs} and high hardware overhead due to the nonzero-index buffer \cite{nctulaisparsecnn, dateloadbalance, hansoneie, cnvlutin, spiking-od-chang}. The overhead could be worse for the proposed RSNN model due to the support of different time steps. 

To avoid these problems while maximizing zero-skipping benefits, this paper proposes reconfigurable zero-skipping hardware with input broadcasting. Our zero-skipping focuses on skipping zero input due to its higher sparsity than weights, which also eases the hardware design. The zero-skipping is reconfigured to fit data flows of different layers and different time steps. The input broadcasting to all PEs ensures all PEs active during nonzero operations, facilitating optimal hardware utilization without load imbalance. The spike input also saves significant routing area compared to multi-bit input in non-SNN models. This also eliminates the need for an additional index buffer, contrary to \cite{nctulaisparsecnn, dateloadbalance, hansoneie, cnvlutin, spiking-od-chang}.

\begin{figure}[t]
	\centering{\includegraphics[width=0.48\textwidth]{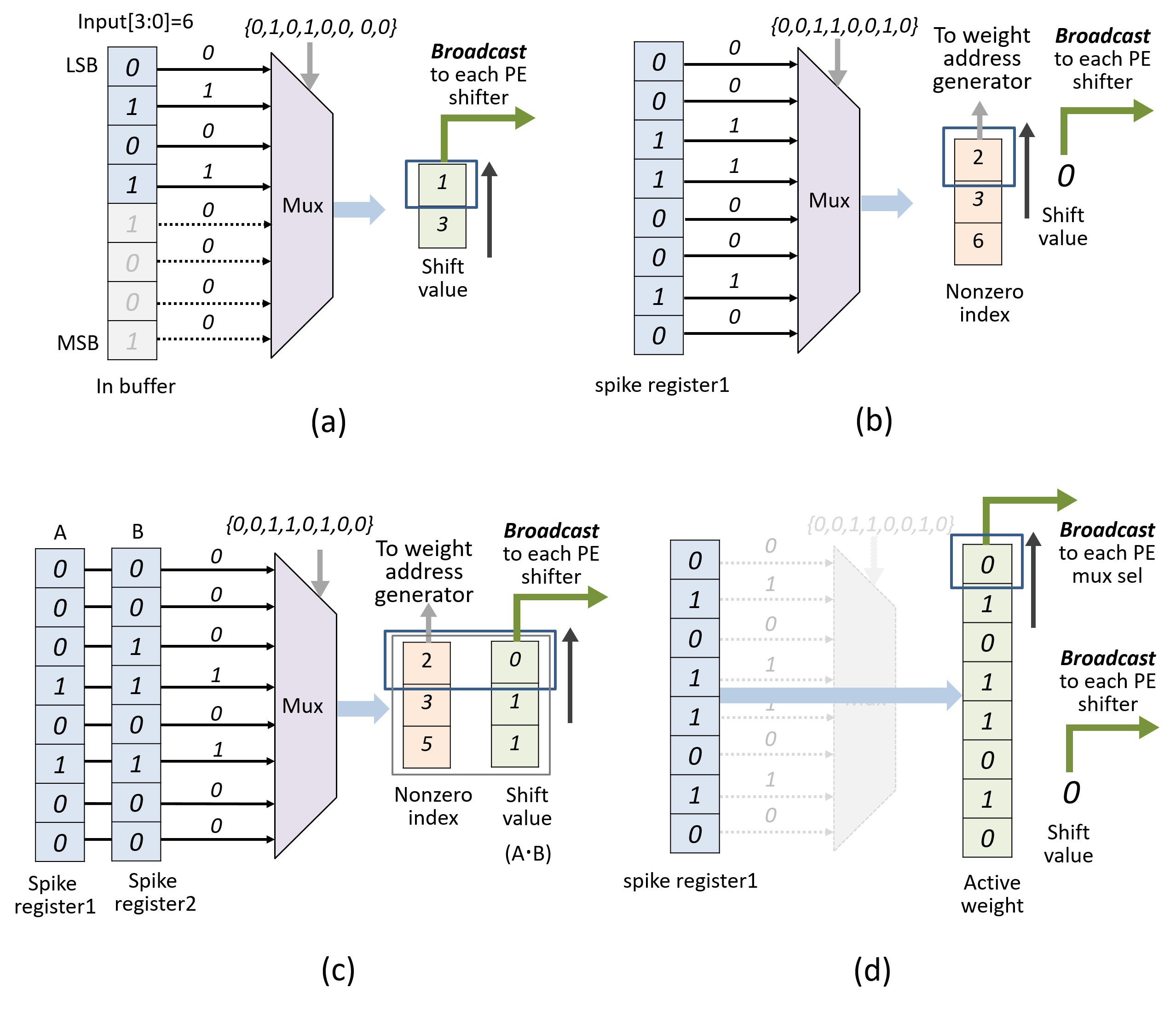}}
        \caption{Reconfigurable zero-skipping: (a) type-A for the input features; (b) type-B for the single time step; (c) type-C: two time steps for the FC layer; (d) type-D: two time steps for the recurrent layer.}
        
	\label{fig:zero-skipping-hw}
\end{figure}

\begin{figure}[t]
	\centering{\includegraphics[width=0.28\textwidth]{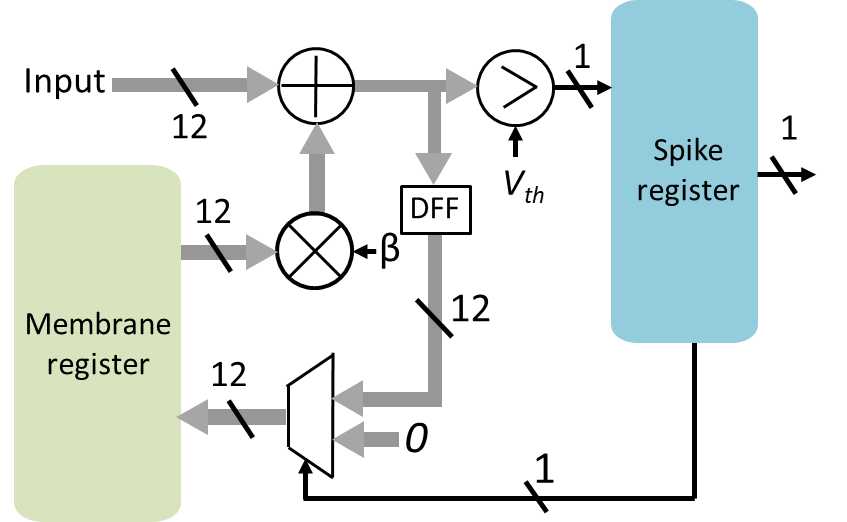}}
	\caption{Leaky Integrate-and-Fire hardware}
	\label{fig:LIF}
\end{figure}

Fig.~\ref{fig:zero-skipping-hw} shows the four data flows for the one set of the zero-skipping unit. The type-A in Fig.~\ref{fig:zero-skipping-hw}(a) is for the input that skips zero input bits. The 8-bit input from the \textit{In Buffer} is processed bit-serially to share the same flow as the spike input in other layers for better hardware utilization. An 8-bit input is split into two 4-bit groups. Each group is assigned to one set of zero-skipping and PEs, enhancing operation speed and PE utilization. For each 4-bit group, the bit index of the nonzero bits is extracted as the left-shift values to the shifter of each PE for the shift-add operation.

The type-B in Fig.~\ref{fig:zero-skipping-hw}(b) is for the single time-step execution in recurrent or FC layers. In this type, the 8 spike signals are read from the \textit{spike register 1} and the corresponding nonzero indexes are extracted one by one to load the corresponding weights from the weight buffer. In this case, the shift value is set to zero. 

The type-C in Fig.~\ref{fig:zero-skipping-hw}(c) is for the two time-step execution in FC layer. This type supports the merged spike by reading two sets of the 8 spike signals: \textit{A} and \textit{B}. These two sets undergo bit-wise \textit{AND} and bit-wise \textit{OR} operations to merge the spikes from two time steps and extract the resulting nonzero index. The corresponding shift value is obtained by \textit{A AND B} for each PE (i.e., left-shift by 1 or 0 bit).

The type-D in Fig.~\ref{fig:zero-skipping-hw}(d) is for the two time-step execution in recurrent layers. This case utilizes \textit{parallel time steps} to reduce weight accesses, and does not skip the zero spikes to avoid using dual-port SRAM and save cost, as explained in Section~\ref{subsubsection:spike_operation}. Thus, its value is broadcast to the PEs directly.

\subsection{Leaky Integrate-and-Fire Hardware}

 Fig.~\ref{fig:LIF} shows the LIF neurons following Eq.~\eqref{eq:2}. The leakage factor \(\beta\) and threshold \(V_{th}\) are set to an approximate power of 2 value to simplify calculations during inference. The adder adds the input stimulus signal with the prior membrane potential, adjusted by \(\beta\), determining the subsequent membrane value. A comparator compares this outcome with \(V_{th}\), generates a spike signal, and stores it in the spike register. The multiplexer then decides whether the membrane value should be reset on the basis of the current spike signal.

\begin{figure}[t]
	\centering{\includegraphics[width=0.39\textwidth]{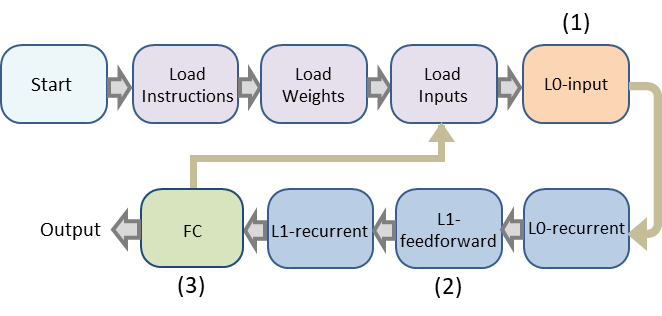}}
	\caption{Finite State Machine for RSNN operations}
	\label{fig:fsm}
\end{figure}

\subsection{Data Flow for Executing the Proposed RSNN Model}

Fig.~\ref{fig:fsm} shows the state machine for the RSNN model. This design first initializes all the required settings by loading instructions, weights, and first input. Then, with \textit{parallel time steps}, it then processes all layers sequentially as listed in Table~\ref{table:baseline}. The details of the RNN (L0-input, L0-recurrent, L1-feedforward, and L1-recurrent) and FC layers are described below.

\subsubsection{Data flow for the L0-input}
\label{subsubsection:input_operation}

\begin{figure}[t]
	\centering{\includegraphics[width=0.43\textwidth]{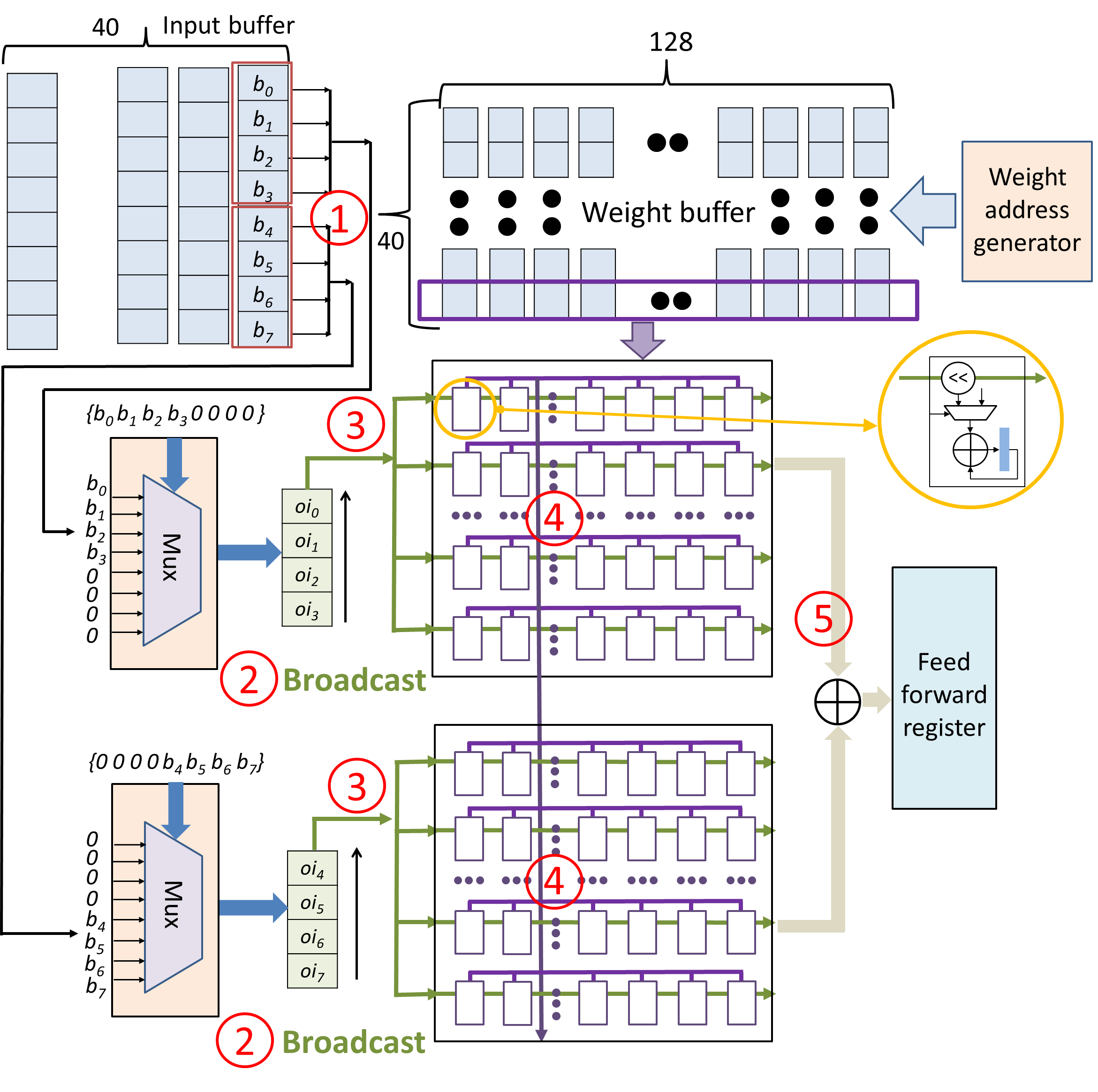}}
	\caption{Data flow for computing the input feature within the DLA}
	\label{fig:dla-input}
\end{figure}

The model input is an 8-bit data. To share and reuse the PEs for spike computation, the 8-bit data is processed bit-by-bit with a shift-and-add method. Fig.~\ref{fig:dla-input} depicts this data flow, which proceeds in five steps.
(1) Split the 8-bit input into two 4-bit groups, and send one group to one set of zero-skipping unit and PEs for speedup. (2)(3) The zero skipping unit uses the type-A configuration to skip zero bit and broadcast the shift-value to all PEs. (4) Each PE selects the shifted weight and accumulates it locally. The 128 weights are loaded from the weight buffer and sent to all PEs. (5) the outputs from both sets of 128 PEs are added to yield the final 128 results, which are then saved in the feedforward register. Subsequently, this result is reused for all time steps to reduce complexity.

\begin{figure}[t]
	\centering{\includegraphics[width=0.44\textwidth]{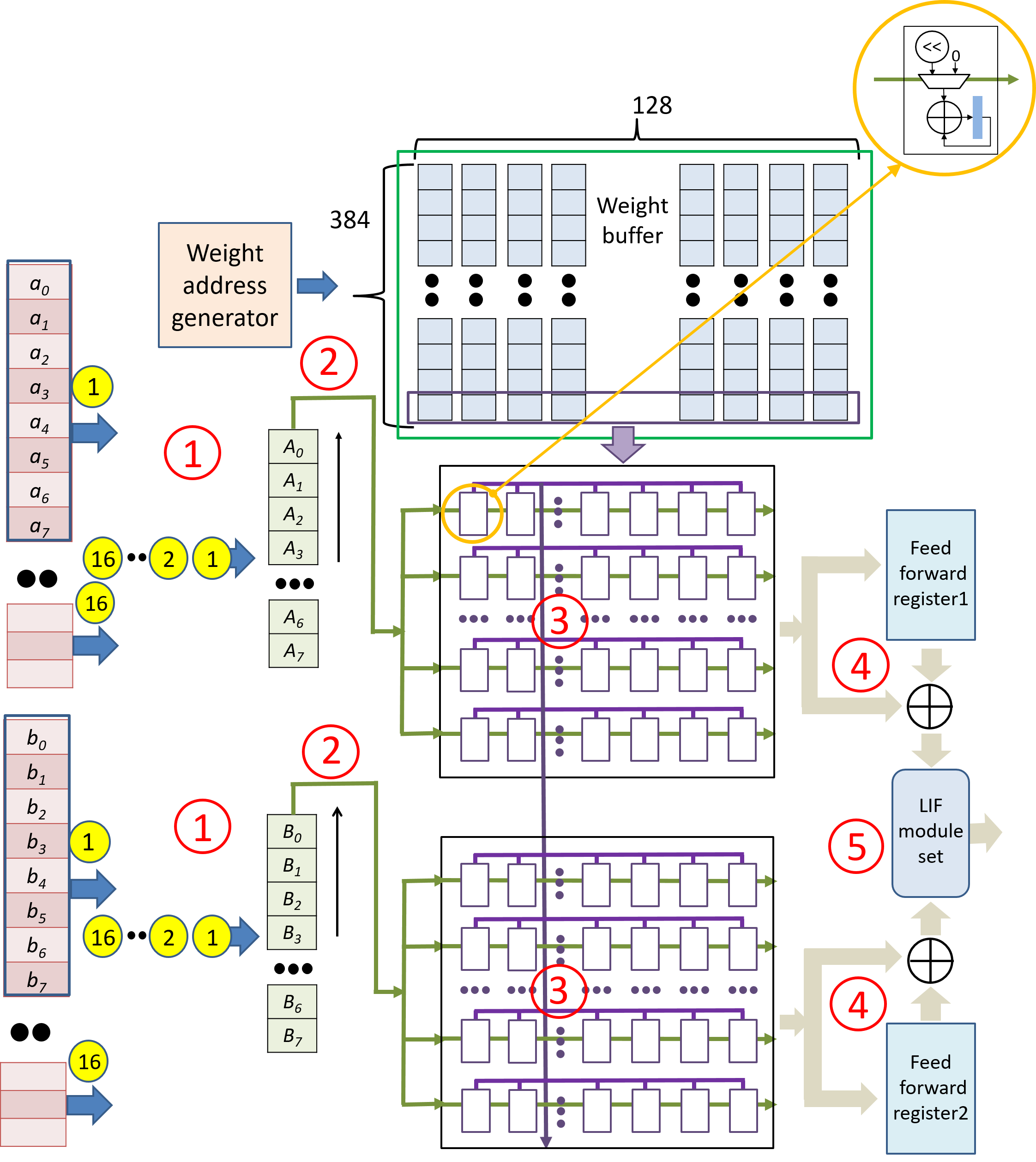}}
	\caption{Data flow for spike computation over two time steps within the DLA}
	\label{fig:dla_spike}
\end{figure}

\begin{figure}[t]
	\centering{\includegraphics[width=0.485\textwidth]{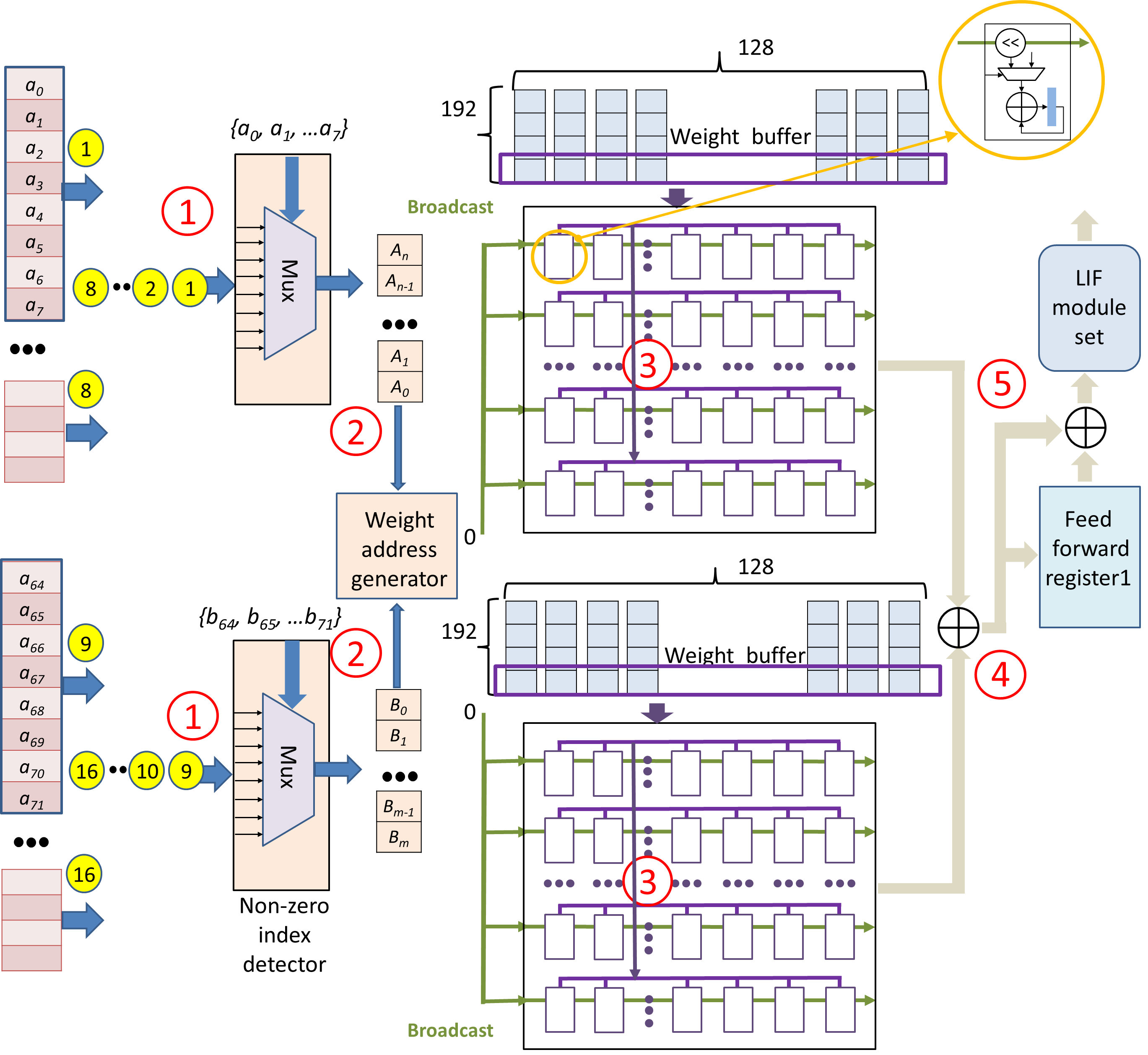}}
	\caption{Data flow for spike computation over single time step in the DLA}
	\label{fig:dla_spike-t1}
\end{figure}

\subsubsection{Data flow of the L0-recurrent, L1-feedforward, and L1-recurrent}
\label{subsubsection:spike_operation}

Utilizing the proposed \textit{parallel time steps} in the DLA data flow allows weights to be shared over two different time steps. This strategy reduces weight access frequency, leading to an energy-efficient design.
Fig.~\ref{fig:dla_spike} shows the data flow for two time steps. In this flow, operations for each time step are allocated to one set of PEs. The two sets of PEs can support two time-step at the same time, which is the proposed \textit{parallel time steps}. This flow includes five steps.
(1) Two groups of 128 spike input are divided into 16 sets of 8 spike input each. 
(2) Load two sets of 8 spike input for each PE set and store them in spike registers as the type-D. 
(3) The spike inputs are processed one by one and broadcast to all corresponding PEs to determine whether the new weight is added to its designated accumulator. The weights are shared for two PE sets. 
(4) After processing the two 128 spike signals, the resulting 128 12-bit values are either stored in the feedforward registers, added with the input operation outcome, or combined with the feedforward register from spike computation results. 
(5) These two sets of 128 12-bit results are forwarded to the LIF modules for the final spike generation. The resulting spike outcomes are stored in the corresponding spike registers for subsequent operations. Note that due to the combined application of \textit{parallel time steps} and \textit{zero-skipping} techniques, dual-port SRAM becomes essential. This configuration consumes roughly three times the area of single-port SRAM. As a consequence, the \textit{zero-skipping} scheme is excluded from this operation.

The operations of the single time step are similar to the above flow with some minor changes. Fig.~\ref{fig:dla_spike-t1} illustrates this single time step process. The major difference is that this flow adopts the type-B zero-skipping with individual weights for each PE set.

\subsubsection{Data flow for the FC layer}
\label{subsubsection:FC_operation}

\begin{figure}[t]
	\centering{\includegraphics[width=0.42\textwidth]{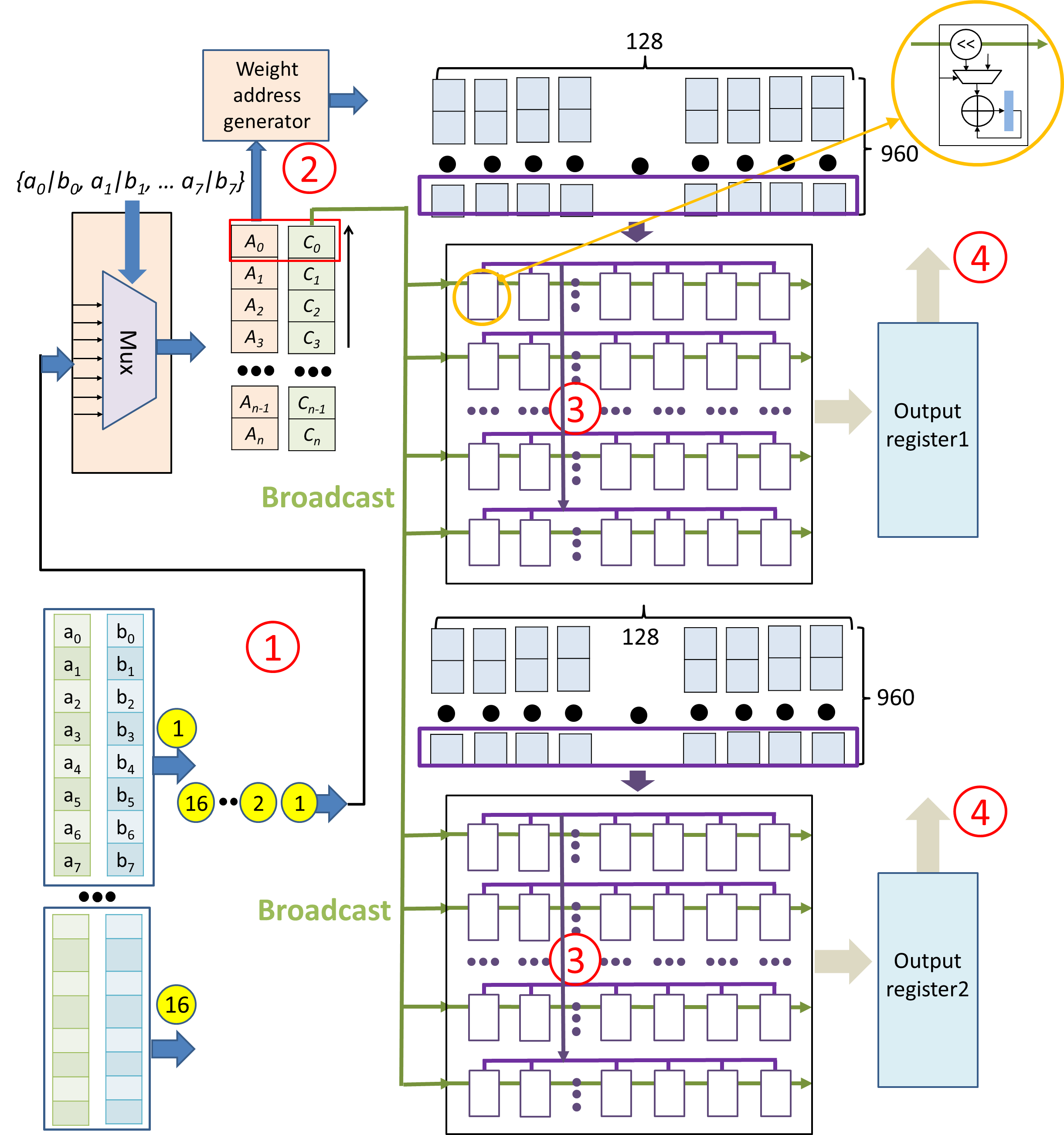}}
	\caption{Data flow of the FC computations within the DLA}
	\label{fig:dla-output1}
\end{figure}

Fig.~\ref{fig:dla-output1}  shows the data flow for the FC layer, including four steps. (1) To support the \textit{merged spike} technique, this design adopts the type-C configuration for input and zero skipping unit. The two 8 spike inputs are processed with bit-wise \textit{AND} and bit-wise \textit{OR} to control input and selection of PE multiplexers. (2) Skip the zero spike. Generate the shift values and broadcast them to all PEs to control weight shift. The generated the 3-bit non-zero index value is sent to the weight address generator to fetch the corresponding weights of the nonzero spike. (3) PEs accumulate the shifted weights until all merged spike operations are completed. (4) The accumulation results are sequentially shifted to the output in groups of 4 12-bit values. Note that for the concluding $128 \times 128$ FC calculations, the operations of 128 merged spike signals are divided into two groups to optimize the utilization of both PE sets.

For single time-step executions in the FC layer, the type-B zero-skipping hardware is employed. The PE operation flow is similar to  FC operations in two time steps. Details are omitted here for clarity.

\begin{figure}[t]
	\centering{\includegraphics[width=0.45\textwidth]{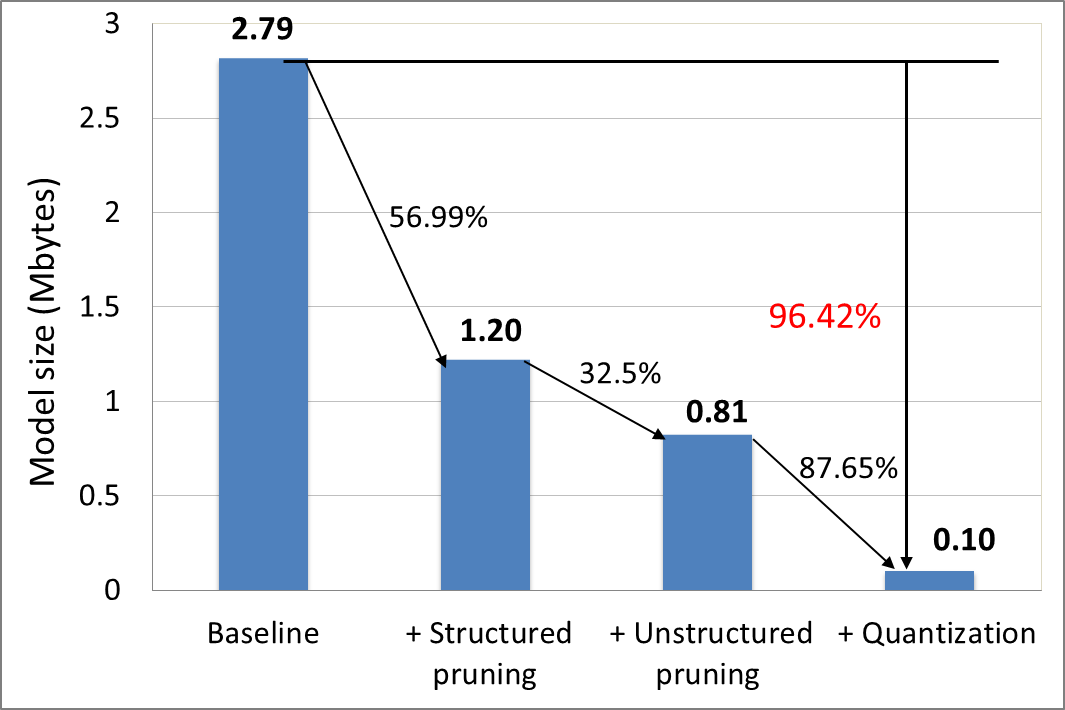}}
	\caption{Weight size reduction with various model compression techniques}
	\label{fig:weight-reduction}
\end{figure}

\begin{figure}[t]
	\centering{\includegraphics[width=0.48\textwidth]{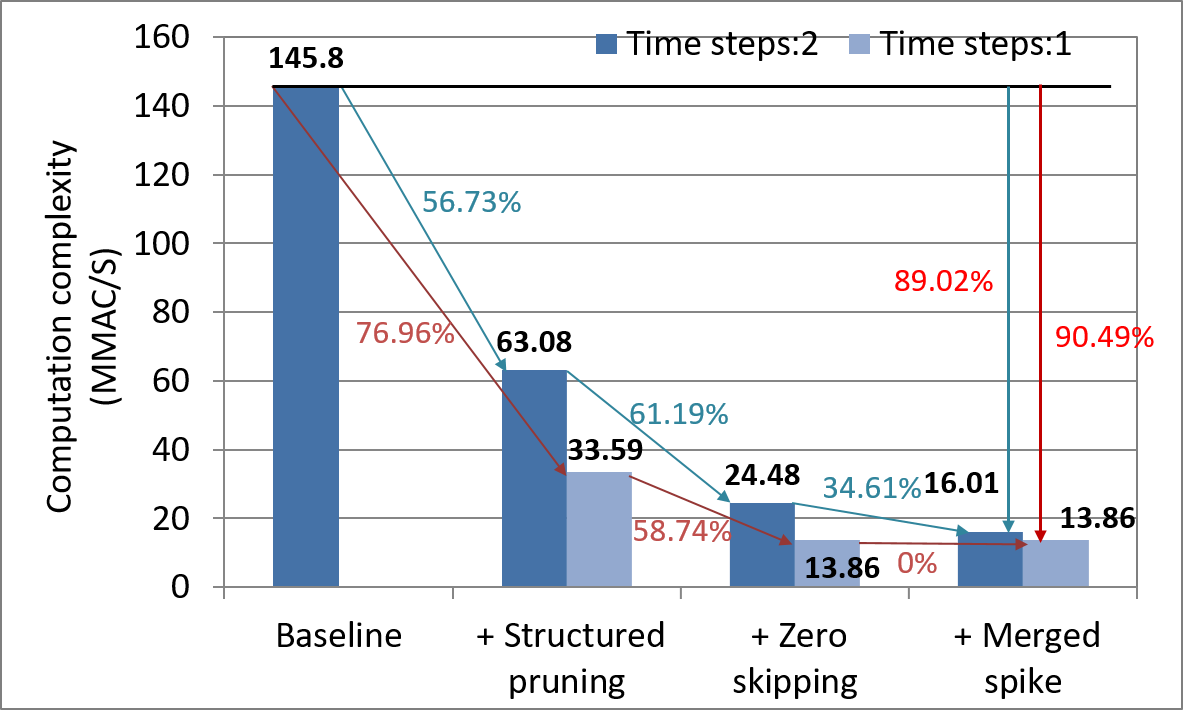}}
	\caption{Computational complexity using various techniques (Baseline is with two time steps)}
	\label{fig:computation-comparison}
\end{figure}

\section{Experimental Results}
\subsection{Model Training}

The proposed model is implemented in PyTorch and evaluated on the TIMIT dataset \cite{timit} using the PyTorch-Kaldi ASR toolkit \cite{pytorch} for feature extraction preprocessing and decoding, a renowned tool extensively employed in various ASR research works \cite{jsscc-lstm-asr, wu2023deep-1, wu2020deep-2, spiking-lstm-energyefficient2022, spiking-lstm-mutli-bit-2022}. During training, the RSNN model utilizes the Adam optimizer for a span of 72 epochs, beginning with a learning rate of $3.5e^{-3}$ which decreases by $1e^{-3}$ every 24 epochs. The loss function is cross entropy. For comparative analysis, we use the RSNN model discussed in Section \ref{subsection:baseline_model} as our baseline for comparison with compressed models.

\subsection{Model Size and Computation Complexity}

Fig.~\ref{fig:weight-reduction} and Fig.~\ref{fig:computation-comparison} depict the reductions in model size and computational complexity achieved through various optimization techniques. As mentioned in Section~\ref{subsection:baseline_model}, the baseline model is characterized by recurrent layer dimensions of 256 and FC dimensions of 1920, and occupies a size of 2.79 MB in a 32-bit floating-number format. After applying structured pruning, the recurrent layer dimensions are reduced to 128 while the FC dimensions remain at 1920, resulting in a 1.20 MB model size or a model size reduction of 56.99\%. Unstructured pruning further reduces the weight in the FC layer by 40\%, resulting in a 0.81 MB model size or a model size reduction of 32.5\%. Table~\ref{table:baseline} provides detailed dimensions for layers in the compressed model. Additionally, using the quantization-aware training method, which quantizes the data format from a 32-bit float to a 4-bit fixed point, further compresses the model to 0.1 MB, a notable reduction of 87.65\%. When combined, the techniques lead to a reduction of 96.42\% in the model's size.

Regarding computational complexity with two time steps, Fig.~\ref{fig:computation-comparison} highlights the savings in computation operations: from an initial 145.8 MMAC/S to 63.08 MMAC/S and 24.48 MMAC/S after structured pruning and sparse activation zero-skipping, achieving reductions of 56.67\% and 61.79\%, respectively. The \textit{parallel time steps} target weight access reduction and weight buffer cost alleviation, bypassing zero-skipping hardware during spike computations in the initial recurrent layers. By adopting the \textit{merged spike} technique, the computations drop further to 16.01 MMAC/S, a reduction of 34.61\%. For single-step operations, computations fall to 33.59 MMAC/S and 13.86 MMAC/S after structured pruning and \textit{zero-skipping}, translating into reductions of 76.96\% and 58.74\%. In essence, the total computations have been reduced by  89.02\% and 90.49\% for two and one time step, respectively.

\begin{figure}[t]
	\centering{\includegraphics[width=0.48\textwidth]{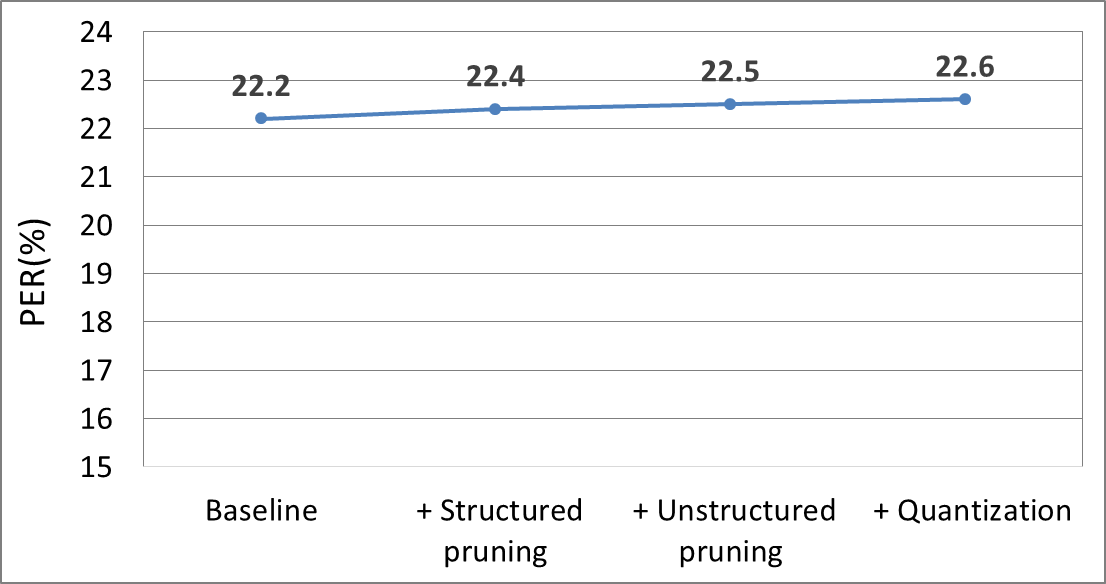}}
	\caption{Error rate evaluated with various model compression techniques}
	\label{fig:test-accuracy}
\end{figure}

\begin{figure}[t]
	\centering{\includegraphics[width=0.48\textwidth]{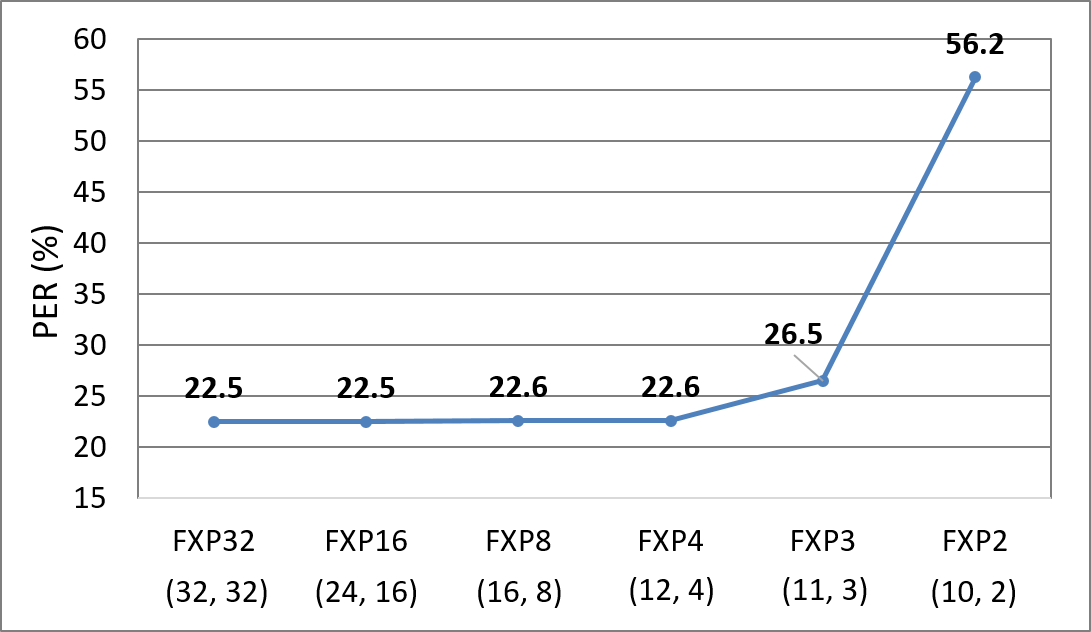}}
	\caption{Error rate across different quantization data widths using the pruned network model (dimension=128). Note: In the format (m, n), m indicates the number of bits for membrane potential, and n represents the number of bits for weight.}
	\label{fig:quant-result}
\end{figure}

\begin{figure}[t]
	\centering{\includegraphics[width=0.48\textwidth]{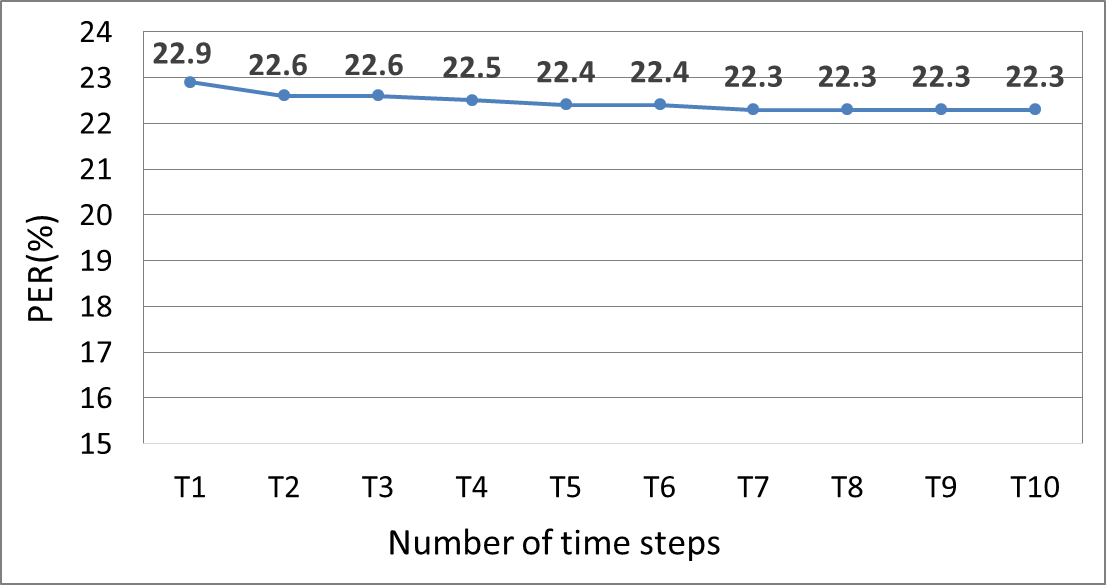}}
	\caption{Error rate evaluated with various number of time steps}
	\label{fig:time-steps-test-accuracy}
\end{figure}

\begin{figure}[t]
	\centering{\includegraphics[width=0.48\textwidth]{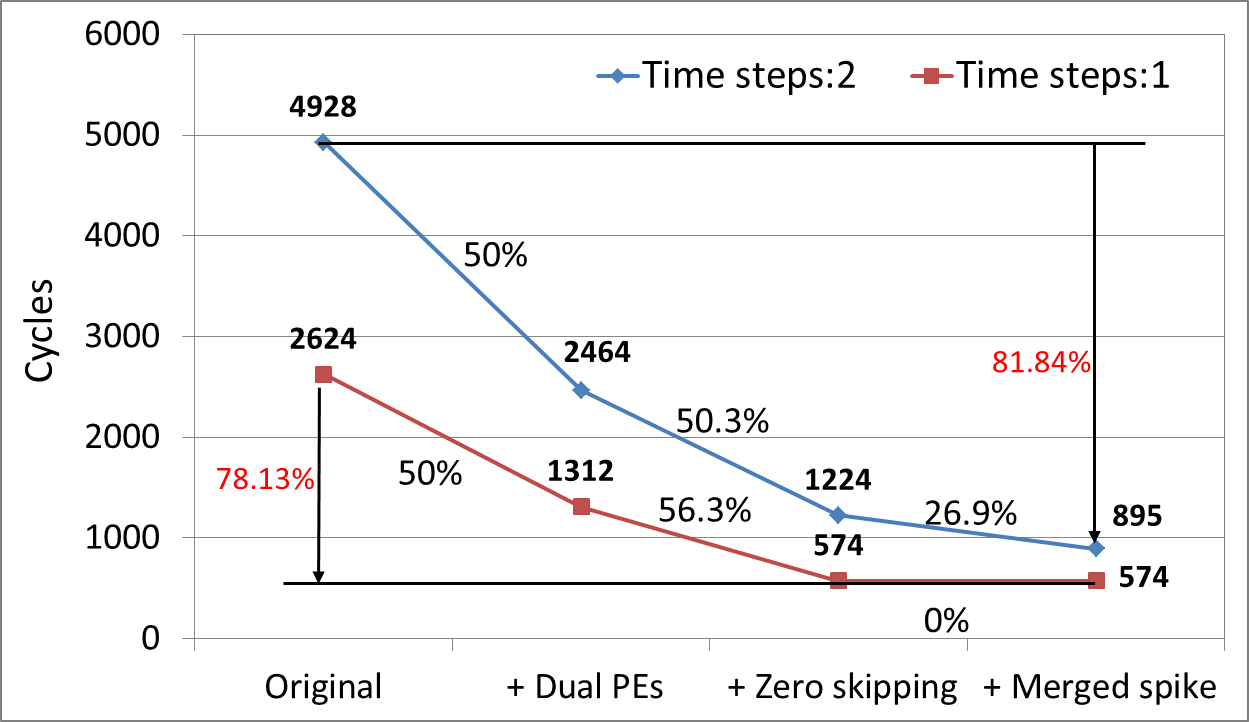}}
	\caption{Cycle count for one and two time steps when executing a single frame of data}
	\label{fig:cycle-count-one-frame}
\end{figure}

\begin{figure}[t]
	\centering{\includegraphics[width=0.45\textwidth]{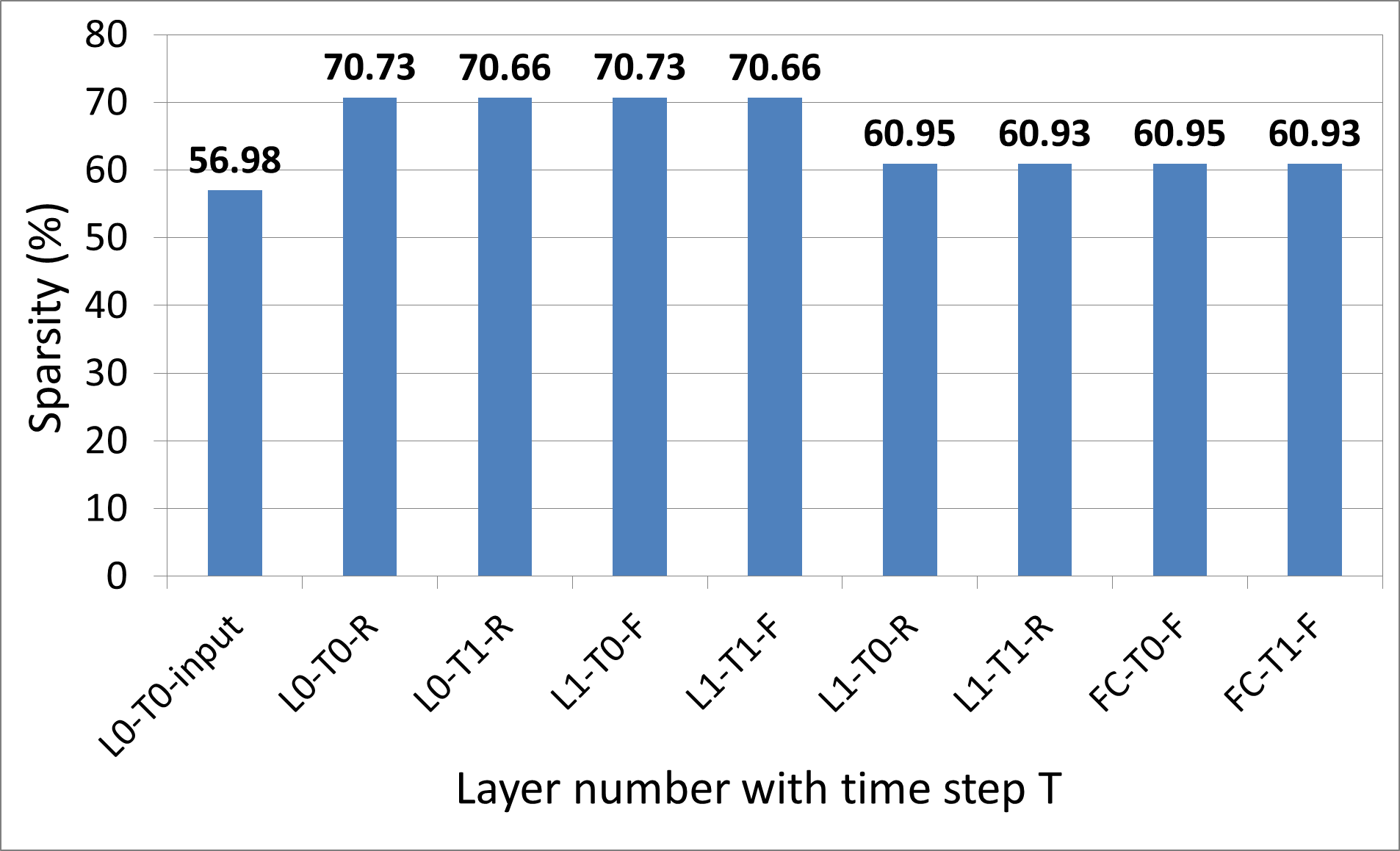}}
	\caption{Sparsity across each layer and time step. Note: $L$ represents the layer number; $T$ denotes the time step number; $R$ symbolizes the recurrent connection, and $F$ signifies the feedforward connection.}
	\label{fig:sparsity}
\end{figure}

\subsection{Ablation Study}

 Fig. 14 illustrates the impact of the proposed optimization techniques on the phoneme error rate (PER). The PER experiences a negligible increase, rising from 22.2\% (baseline) to 22.4\% (+structured pruning), 22.5\% (+unstructured pruning), and finally to 22.6\% (+quantization).

Fig.~\ref{fig:quant-result} depicts the quantization impacts of various formats. The 4-bit fixed point format maintains performance closely matching the 32-bit floating point. Thus, the 4-bit fixed point format is selected for weight values, resulting in a 87.5\% reduction in the baseline model size.

Fig.~\ref{fig:time-steps-test-accuracy} illustrates the error rate evaluated over varying numbers of time steps. An increase in time steps led to a marginal enhancement in error rate, attributed to the recurrent neural network's temporal information capture capability. Consequently, our model employs one or two time steps for the speech recognition task.

Fig.~\ref{fig:cycle-count-one-frame} depicts the cycle count reduction achieved through the proposed optimization techniques. With dual sets of PEs, cycle counts for one and two time steps are 1312 and 2464, respectively, a reduction 50\%. The \textit{zero-skipping} technique further reduces these to 574 and 1224 cycles, a decrease of 56.3\% and 50.3\% respectively. Using the \textit{merged spike} technique for two time steps yields a count of 895, a 26.9\% reduction. Overall, the reductions amount to 78.13\% and 81.84\% for one and two time steps, respectively.

Fig.~\ref{fig:sparsity} depicts sparsity across each layer and time step with two time steps. The input bits exhibit approximately 57\% sparsity, while other layers and time steps vary between 60\% and 71\%. Due to this pronounced sparsity, zero-spike operations can be bypassed with our zero-skipping hardware, leading to a reduced computation cycle count.

\begin{figure}[t]
	\centering{\includegraphics[height=44mm]{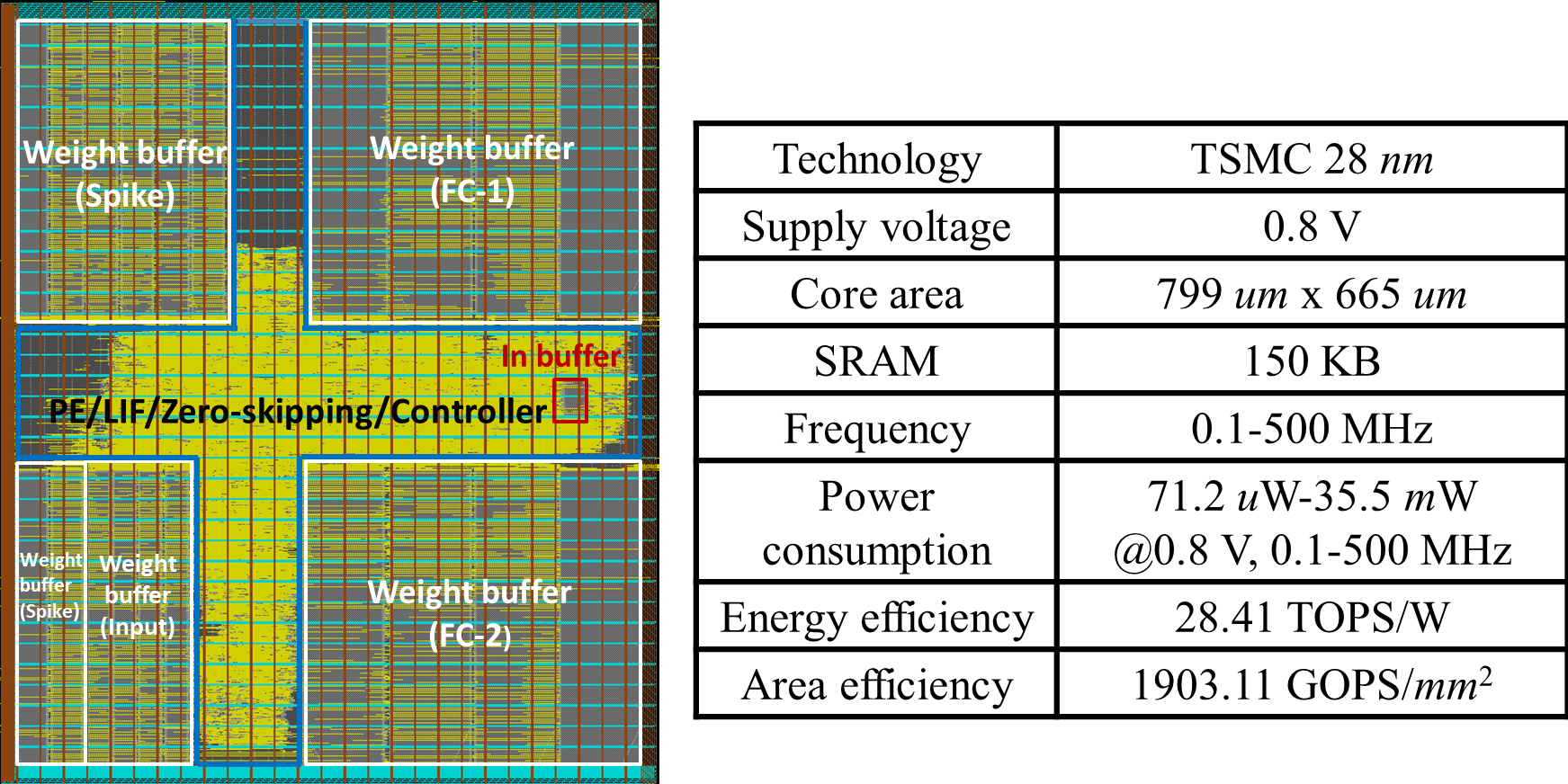}}
	\caption{DLA design layout and performance summary}
	\label{fig:new_chip_photo}
\end{figure}

\begin{figure}[t]
	\centering{\includegraphics[width=0.48\textwidth]{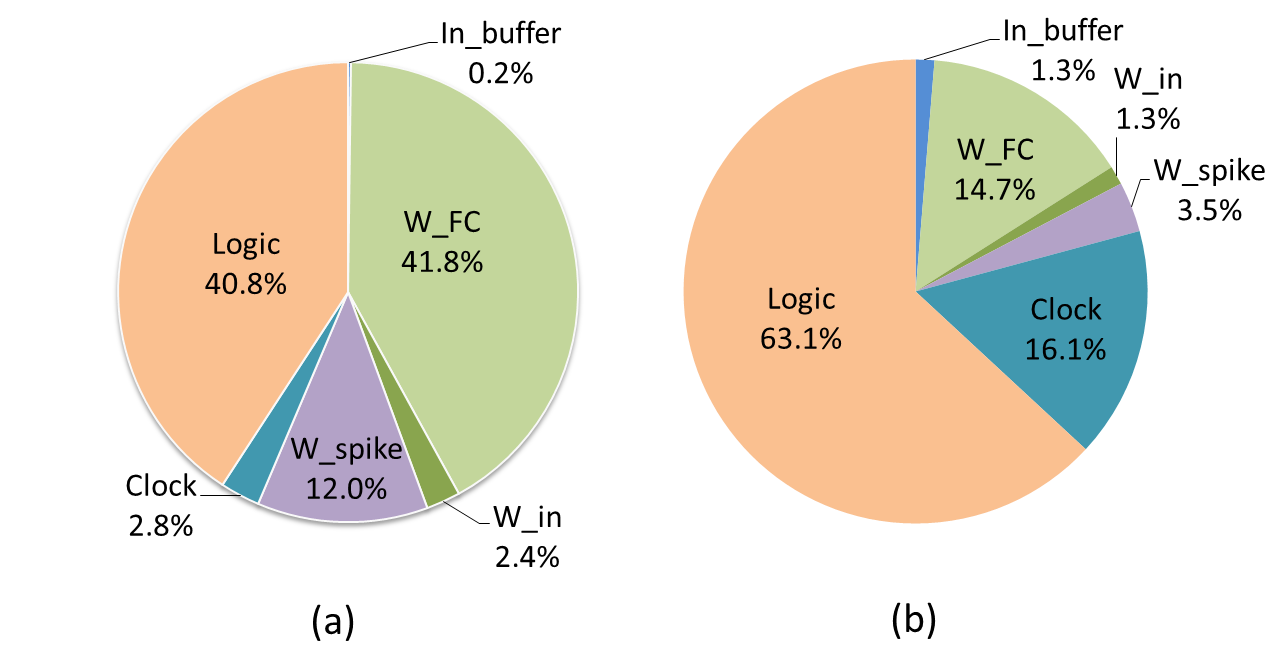}}
	\caption{Power breakdown of the DLA design: (a) at 100 kHz; (b) at 500 MHz }
	\label{fig:chip-power-breakdown}
\end{figure}

\begin{table}[]
\centering
\caption{Comparisons with model in \cite{spiking-lstm-mutli-bit-2022}}
\label{table:model_comparision}
\begin{tabular}{|c|c|c|c|c|c|}
\hline
  & \begin{tabular}[c]{@{}c@{}}Recurrent\\ SNN \end{tabular} & \begin{tabular}[c]{@{}c@{}} SNN \\ time-step\end{tabular} & \begin{tabular}[c]{@{}c@{}}Learnable\\ \(V_{th}\)/decay\end{tabular} & \begin{tabular}[c]{@{}c@{}}Inherent\\ temporal\\ training \end{tabular} & \begin{tabular}[c]{@{}c@{}}Output\end{tabular} \\ \hline

\cite{spiking-lstm-mutli-bit-2022} & \checkmark &   &   &    & 6-bit \\ \hline

This Work& \checkmark    & \checkmark   & \checkmark & \checkmark  & Spike     \\ \hline
\end{tabular}
\end{table}

\begin{table*}[t]
	\centering
	\caption{Comparisons with other DLA designs}
	\label{table:overview}
	\begin{tabular}{|l||c|c|c|c|c|c|c|}

\hline
\multicolumn{1}{|c|}{}  &  
\multicolumn{1}{c|}{TCAS-II 2022 \cite{ntu-timit}} & 
\multicolumn{1}{c|}{JSSC 2020 \cite{jsscc-lstm-asr}}& 
\multicolumn{1}{c|}{{$^c$}JSSC 2023 \cite{jsscc-denoise-asr2023}} &
\multicolumn{1}{c|}{TCAS-I 2019 \cite{tcas-1-bcnn}} &
\multicolumn{1}{c|}{This Work} \\ \hline 

Technology       & 28\ nm      & 65\ nm  & 16\ nm & 28\ nm & 28\ nm     \\ \hline
Supply voltage    & 0.57-0.9 V  & 0.68-1.1 V & 0.55 V & 0.57-0.9V  & 0.8 V    \\ \hline
Application       &Speech Recognition &Speech Recognition &Speech Recognition  &Speech Recognition &Speech Recognition     \\ \hline
Algorithm  & RNN  & {$^d$}LSTM & RNN & Binary Weight CNN & Recurrent SNN     \\ \hline
Dataset  & TIMIT  & TIMIT & LibriSpeech & TIMIT & TIMIT     \\ \hline
Precision  & {W(FXP 12b)}  & W(FXP 6b)/A(FXP 13b) & W(FP 8b)/A(FP 8b) & A(1b/16b)/W(1b) & A(1b)/W(FXP 4b)      \\ \hline
PE operation type  & MAC  & MAC  & MAC & Adder  & Adder     \\ \hline
No. of PEs  & 16  &  65  & 4 & N/A & 256     \\ \hline
Sparse processing     & Yes (Weight)         & No        & No & No   & Yes (Activation)             \\ \hline
Frequency (MHz)    & 100         & 8-80   & 573   & 2.5-50    & 0.1-500               \\ \hline
SRAM     & 248 KB           &  297 KB   & 5.03 MB  &  52 KB    & 150 KB   \\ \hline
Logic gates     & 1658k           &  N/A   & N/A  &  780k    & 174k   \\ \hline
Latency     & 0.7 ms           &  N/A   &  45 ms &  0.5 ms-10 ms   &  1.79 us-8.95 ms   \\ \hline
Error rate     & 22.8\%           &  20.6\%   &  10.54\% & {$^{e}$}21.58\%;{$^{f}$}25.2\%    &  22.6\%   \\ \hline
Core area  (mm$^2$)    & 2.79      & 4.97   & 8.84 & 1.29   & \textbf{0.531}    \\ \hline
Core power   & 1.7 mW    & 1.85 mW-67.3 mW  & 19 mW  & 141 uW-2.85mW    & \textbf{71.2 uW}-35.5 mW           \\ \hline

\multirow{2}{3.3cm}{Peak core energy efficiency (TOPS/W)}   &2.71  & {$^d$}8.93 & {$^c$}7.8 & 90 & 28.41 \\
  &{$^b$}3.43    & {$^{bd}$}14.98  & {$^{bc}$}2.11 & {$^b$}48.95 &  28.41    \\ \hline 
  
\multirow{2}{3.3cm}{Peak core area efficiency (GOPS/mm$^2$)}  & 2.79  & {$^d$}120.9   & {$^c$}16.76  & 9.91  & \textbf{1903.11}  \\
  & 2.79   &  {$^{ad}$}280.7 &  {$^{ac}$}6.71 & 9.91    & \textbf{1903.11}    \\ \hline

\multirow{2}{3.3cm}{Energy per frame (nJ/frame)}  & 2429  & N/A   & 855k & 141  & \textbf{63.5}  \\
  & {$^{g}$}1919   &  N/A &  {$^{g}$}3165k & {$^{g}$}277.7    & \textbf{63.5}    \\ \hline

		\multicolumn{2}	{|l}{$^{a}$Technology scaling ($\dfrac{process}{28\ nm}$).} & \multicolumn{4}	{l|}{$^{b}$Normalized~energy~efficiency $=$ energy~efficiency $\times (\dfrac{process}{28\ nm}) \times (\dfrac{voltage}{0.8\ V})^2$.} \\ 
        \multicolumn{1}	{|l}{$^{c}$FlexASR only.} & 
        \multicolumn{5}	{l|}{{$^{d}$}Not including the FC layer. {$^{e}$}With self-learning.  {$^{f}$}Without self-learning. Best results are highlighted in bold font. }      
         \\

         \multicolumn{5}	{|l}{$^{g}$Normalized~energy per frame $=$ energy per frame$ \times (\dfrac{28\ nm}{process}) \times (\dfrac{0.8\ V}{voltage})^2$. } & \multicolumn{1}	{l|}{} \\ 
         \hline	

\end{tabular}
\end{table*}

\subsection{Hardware Implementation Results}

Fig.~\ref{fig:new_chip_photo} showcases the design layout and performance summary. Implemented using the TSMC 28-$nm$ CMOS process, the accelerator spans a core area of $799\ \mu m \times 665\ \mu m$ and includes a 150-KB SRAM. For layout area distribution, the input weight buffer, spike weight buffer, FC weight buffer, in-buffer, and PE logic occupy 6.91\%, 18.31\%, 42.84\%, 0.29\%, and 31.65\% of the total area, respectively.

For the pruned speech recognition model with 25-$ms$ speech features, the latency is 574 clock cycles for a single time step and 895 for two. Hence, at a 100 kHz operating frequency, processing times are 5.74 $ms$ and 8.95 $ms$ for one and two time steps, respectively. This confirms the accelerator's real-time speech recognition capability.

Post-layout simulation and power analysis indicate a power consumption of 71.2 $\mu$W at 100 kHz and 0.8 V supply voltage for the two-time step scenario. At 500 MHz, the design achieves an energy efficiency of 28.41 TOPS/W and an area efficiency of 1903.11 GOPS/mm$^2$, credited to the RSNN model adoption, algorithm and architectural optimizations, and gating of idle logic and buffers.

Fig.~\ref{fig:chip-power-breakdown} presents the power breakdown of the DLA implementation at frequencies of 100 kHz and 500 MHz. At 100 kHz, the memory buffer has the highest power consumption at 56.4\%, whereas the logic circuits and clock buffer account for 40.8\% and 2.8\%, respectively. On the other hand, at 500 MHz, the logic circuits demand the most power, representing 63.1\%, while the memory buffer and clock buffer consume 20.8\% and 16.1\%, respectively. By applying the proposed \textit{parallel time steps} technique, the weight access count in the recurrent and FC layers is reduced to around 50\%.

\subsection{Design Comparisons}

Table ~\ref{table:model_comparision} compares our proposed model with the RSNN model in \cite{spiking-lstm-mutli-bit-2022}. Our work possesses several advantages. It supports recurrent SNN time steps, dynamically adjusts the neuron firing threshold, and the decay factor via learnable \(V_{th}\) and decay during training, and captures inherent temporal patterns efficiently. Thus, our model can operate with a low time-step and single-bit spike output. In contrast, the model in [15] lacks these enhancements; as a result, it must utilize multibit output and a larger model size to maintain comparable accuracy.

Table~\ref{table:overview} compares our speech recognition design with others. While designs in \cite{ntu-timit, jsscc-lstm-asr,jsscc-denoise-asr2023, tcas-1-bcnn} utilize ANN models, our design uniquely implements the RSNN model. Comparisons are challenging due to different design operation conditions and various design architectures. 

Our RSNN accelerator, as detailed in Table~\ref{table:overview}, demonstrates the lowest power consumption (71.2 $\mu$W), the smallest design footprint (0.531 mm$^2$), the highest area efficiency (1903.11 GOPS/mm$^2$), and the lowest energy per frame (63.5 nJ/frame) among its counterparts. These four standout results are emphasized in bold within Table~\ref{table:overview}. Such superior performance is attributed to the implementation of the RSNN network model, augmented by optimizations at both the algorithmic and hardware levels, as elaborated in Sections II and III. 

In the context of the studies \cite{ntu-timit} and \cite{jsscc-lstm-asr}, the RNN-based attention accelerator and the LSTM-based accelerator both exhibit higher power consumption, as well as lower normalized peak energy efficiency and area efficiency, compared to our RSNN accelerator. In the \cite{jsscc-denoise-asr2023} study, the RNN-based accelerator integrated into an SoC consumes more power than our RSNN accelerator. Additionally, it exhibits longer latency, along with lower normalized peak energy efficiency and area efficiency. The error rate is significantly reduced to 10.54\%, attributed to the use of the LibriSpeech dataset. 

Compared to the binary weight CNN design presented in \cite{tcas-1-bcnn}, our RSNN accelerator excels in core area efficiency but has lower energy efficiency, primarily due to the widespread use of 1-bit adder operations in \cite{tcas-1-bcnn}. By leveraging a self-learning technique, the binary weight CNN design achieves a reduction in the error rate from 25.20\% to 21.58\%. However, it exhibits higher power consumption compared to our RSNN accelerator. For a fair comparison between these two designs, we adopt the energy per frame from \cite{tcas-1-bcnn}. The energy per frame of our RSNN design, at 63.5 nJ/frame, surpasses that of \cite{tcas-1-bcnn}.  

In conclusion, our design's superior power, area, and efficiency attributes stem from the RSNN architecture and the synergistic optimization of both algorithm and hardware.

\section{Conclusions}

This paper proposes a 71.2-$\mu$W speech recognition accelerator designed for real-time and always-on edge devices through algorithm and hardware co-optimization. At the algorithm level, this design proposes a low time-step recurrent spiking neural network to exploit the benefits of spiking computation and high spike sparsity. The model size and computational complexity are also reduced by 96\% and 90\%, respectively, with \textit{mixed-level pruning}, quantization, \textit{zero-skipping}, and \textit{merged spike} techniques. We further tackle the design challenges of multiple time-step execution with \textit{parallel time steps} and \textit{merged spike} techniques to reduce weight buffer access and cycle count. For the hardware, this design adopts a simple zero-skipping mechanism with a broadcasting approach to exploit the high spike sparsity and alleviate load imbalance problems and area overhead for the non-zero-index buffer. When implemented using the TSMC 28-$nm$ process, our design demonstrates real-time speech recognition capabilities with lower power and area. Moreover, when operated at 500 MHz, the accelerator delivers 28.41 TOPS/W energy efficiency and 1903.11 GOPS/mm$^2$ area efficiency, surpassing state-of-the-art designs.

\section*{Acknowledgment}
The authors would like to thank TSRI for supporting EDA design tools.

\bibliographystyle{IEEEtran}
\bibliography{IEEEabrv, thesis}

\begin{IEEEbiography}[{\includegraphics[width=1in,height=1.25in,clip,keepaspectratio]{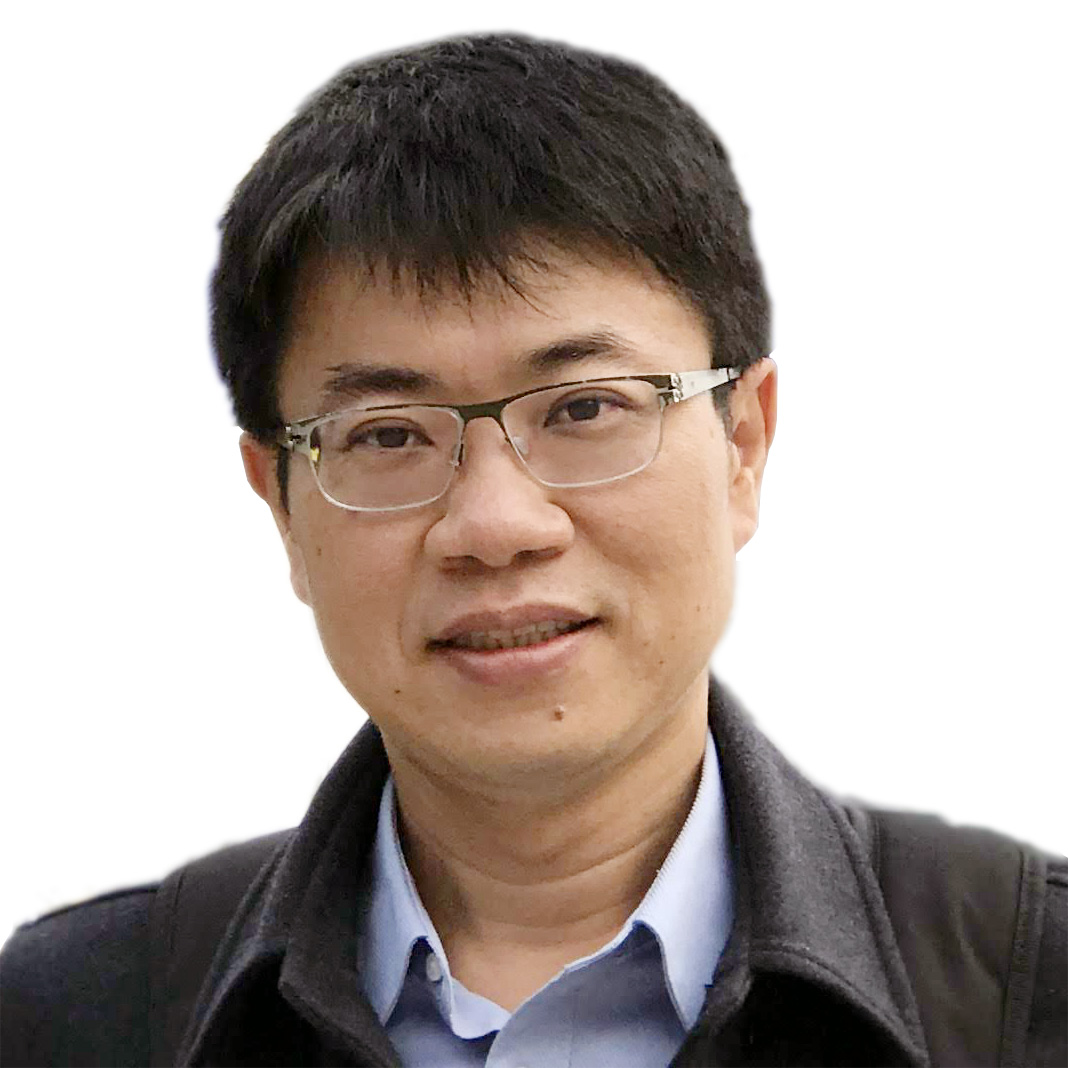}}]{Chih-Chyau Yang}
received the B.S. degree in electrical engineering from National Cheng-Kung University (NCKU), Taiwan in 1996, and the M.S. degree in electronics engineering from National Chiao-Tung University, Hsinchu (NCTU), Taiwan in 1999. He is currently a principal engineer at Taiwan Semiconductor Research Institute (TSRI), Taiwan. His research interests include VLSI design, computer architecture, and platform-based SoC design methodologies.
\end{IEEEbiography}
\begin{IEEEbiography}[{\includegraphics[width=1in,height=1.25in,clip,keepaspectratio]{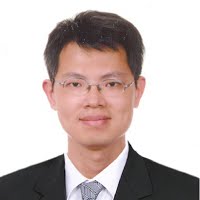}}]{Tian-Sheuan Chang}
	(S’93–M’06–SM’07)
	received the B.S., M.S., and Ph.D. degrees in electronic engineering from National Chiao-Tung University (NCTU), Hsinchu, Taiwan, in 1993, 1995, and 1999, respectively. 
	
	From 2000 to 2004, he was a Deputy Manager with Global Unichip Corporation, Hsinchu, Taiwan. In 2004, he joined the Department of Electronics Engineering, NCTU (as National Yang Ming Chiao Tung University (NYCU) in 2021), where he is currently a Professor. In 2009, he was a visiting scholar in IMEC, Belgium. His current research interests include system-on-a-chip design, VLSI signal processing, and computer architecture.
	
	Dr. Chang has received the Excellent Young Electrical Engineer from Chinese Institute of Electrical Engineering in 2007, and the Outstanding Young Scholar from Taiwan IC Design Society in 2010. He has been actively involved in many international conferences as an organizing committee or technical program committee member.
\end{IEEEbiography}

\end{document}